\documentclass[sn-mathphys-num,Numbered]{sn-jnl}

\usepackage{amsmath,amssymb,amsfonts}
\usepackage{algorithmic}
\usepackage{graphicx}
\usepackage{textcomp}
\usepackage{xcolor}
\usepackage{booktabs}
\usepackage{enumitem}
\usepackage{tabularx}
\usepackage{array}
\usepackage{longtable}
\usepackage{ragged2e}
\newcolumntype{Y}{>{\raggedright\arraybackslash}X}

\usepackage{framed}
\definecolor{shadecolor}{gray}{0.95}
\newenvironment{rqbox}{
  \par\noindent
  \setlength{\FrameRule}{0.4pt}
  \setlength{\FrameSep}{6pt}
  \begin{shaded*}\noindent}{
  \end{shaded*}
}

\begin{document}

\title{An Empirical Study of Complexity, Heterogeneity, and Compliance of GitHub Actions Workflows
}

\author[1]{\fnm{Edward} \sur{Abrokwah}}\email{edwardabrokwah@trentu.ca}

\author[1]{\fnm{Taher A.} \sur{Ghaleb}}\email{taherghaleb@trentu.ca}

\affil[1]{\orgname{Trent University}, \city{Peterborough}, \country{Canada}}

\abstract{
Continuous Integration (CI) has become a core practice in modern software engineering, enabling rapid and collaborative software delivery. GitHub Actions (GHA) has become a leading CI platform due to its tight GitHub integration and growing ecosystem of reusable workflows. Despite extensive documentation and best practices, there is limited empirical understanding of how real-world GHA workflows align with recommended guidelines. This study analyzes the structure, complexity, heterogeneity, and compliance of GHA workflows across Java, Python, and C++ repositories. We (a) quantify workflow complexity, (b) identify recurring and diverse structural patterns, (c) evaluate compliance with best practices, and (d) compare workflow design across languages. GHA workflows are generally small, shallow, and heavily dependent on external actions, with limited sequence-level standardization despite recurring intent-level patterns. A common pipeline prefix appears in 39.5\% of workflows, but workflow sequences are highly heterogeneous, with only one exceeding the 5\% global frequency threshold. We observe that Java has the lowest explicit test adoption, Python follows canonical templates but shows weaker security practices, and C++ workflows are larger and more structurally diverse. Compliance gaps are widespread, especially in permissions, timeout configuration, and SHA pinning, while reusable workflows remain rare. Build-without-test patterns suggest that workflow design is driven more by ecosystem conventions and platform defaults than by best practices. This indicates that better defaults and tooling could improve workflow security, modularity, and maintainability, while researchers should account for language- and repository-level factors when analyzing CI systems. Overall, this work provides a reproducible empirical baseline for studying and improving GHA workflow design in open-source ecosystems.
}

\maketitle

\section{Introduction}
Continuous Integration (CI) is central to modern software development, helping developers automatically test, build, and deploy their code, enabling early detection of integration problems~\cite{Fowler_CI}.
Software repositories normally adopt CI services by writing dedicated YAML configuration files~\cite{valenzuela2024hidden}.
GitHub Actions (GHA) is a GitHub-integrated CI service that has become one of the most widely adopted CI services to run CI workflows directly within GitHub repositories since its introduction in late 2019~\cite{golzadeh2022rise}. This uptake is largely due to GHA's tight integration with the GitHub ecosystem, and its marketplace hosts thousands of reusable action components, enabling developers to incorporate common tasks without writing custom scripts~\cite{wessel2023github}. GHA workflows consist of jobs (which can run sequentially or in parallel), each of which contains ordered steps of commands. This structure enables rapid feedback loops and more consistent deployment pipelines~\cite{valenzuela2024hidden}. However, a recent study found that only about 10\% of GitHub projects include deployment steps in GHA, suggesting that deployment remains underutilized or inconsistently integrated~\cite{ghaleb2025ci_cd_android}. 

CI workflows often involve many jobs, conditional triggers, and integrations with external tools. This complexity can increase the maintenance burden on developers~\cite{valenzuela2024hidden}. Empirical studies have reported that CI workflow files often require frequent updates during development~\cite{valenzuela2024hidden}. To quantify workflow complexity, researchers can measure metrics, such as the number of workflow files or their sizes, which can highlight unusually large or small configurations~\cite{valenzuela2024hidden}. The extent to which workflows reuse common patterns or exhibit diverse structures remains unclear. Understanding this variability is critical, as it can signal a lack of best practices or the emergence of heterogeneity~\cite{github2024}.

Our study investigates structural complexity, usage heterogeneity, and adherence to best practices in GHA workflows across Java, Python, and C++ projects. Using a large dataset of workflows, we assess how configuration size, job structure, and common practices vary both across and within language ecosystems.

Our findings quantify workflow complexity, reveal structural variation, and evaluate alignment with best practices~\cite{ghaleb2025ci_cd_android,kiss2022explorative}. Although earlier research has examined CI complexity in Android projects~\cite{ghaleb2025ci_cd_android} and Travis CI configurations~\cite{gallaba2020tse}, our study is, to our knowledge, the first large-scale analysis to jointly address the complexity, heterogeneity, and compliance with best practices in GHA across programming languages. These insights can inform documentation, tooling, and developer education to improve CI maintainability and adoption.

The remainder of the paper is organized as follows. Section~\ref{sec:Background_and_Related_Work} provides a background context and an overview of the related work.
Section~\ref{sec:Study_Setup} outlines the study setup, including data collection methods, sources, processing, and analysis procedures. Section~\ref{sec:Evaluation} details the motivation, approach, and results for each research question. Section~\ref{sec:discussion} discusses the implications of our findings for CI service providers, researchers, and developers.  Section~\ref{sec:Threats_to_Validity} discusses potential threats to validity and their implications for our findings. Finally, Section~\ref{sec:Conclusion} summarizes the contributions and implications of the study.

\section{Background and Related Work}
\label{sec:Background_and_Related_Work}

\subsection{Background}

\subsubsection{GitHub Actions (GHA) Workflows and Best Practices}
GitHub's official guides emphasize maintainable and reusable workflow design. For example, instead of duplicating logic, developers are encouraged to create reusable workflows to be invoked by others; this "avoids duplication" and "promotes best practice" by reusing well-designed, proven workflow components~\cite{github2024}. 
GHA also provides a marketplace of pre-built actions (for tasks like checkout, setup of languages, etc.), which ideally should be leveraged to follow typical practices. These recommendations form a baseline for how GHA pipelines should be structured~\cite{saroar2023developers}.

\vspace{4pt}
\subsubsection{CI Principles}

CI emphasizes automated testing, early quality checks, and fast feedback to developers. In practice, this translates to pipelines that automatically run tests, perform static and security analyses, and produce deployable artifacts after each commit or pull request. However, open-source projects vary in how comprehensively they apply these principles. While many use CI for building and testing, fewer automate deployment, often due to infrastructure limitations or the nature of the project (e.g., libraries)~\cite{gallaba2020tse}. Though GitHub provides built-in security tools such as CodeQL, it is unclear how widely such security workflows are adopted in practice.

Following CI best practices improves automation, reduces errors, and ensures consistent delivery~\cite{github2024}. If many GHA workflows lack testing or deployment stages, this may indicate missed opportunities to strengthen CI quality. It could also point to gaps in available templates or documentation that support developers in adopting more standardized, effective CI configurations. Even though prior work has examined CI usage variability and common workflow issues~\cite{faqih2024empirical,khatami2024catching,ghaleb2025ci_cd_android}, the extent to which GHA workflows are consistently adopted in practice and whether they reflect established CI practices remains underexplored.

Our work bridges this gap by analyzing workflow complexity and heterogeneity across popular programming languages (Java, Python, and C++), while also assessing the presence or absence of key CI practices. We aim to link low-level CI configuration patterns with higher-level process goals.

\subsection{Related Work}

Several studies have investigated CI workflows across services. Ghaleb et al.~\cite{ghaleb2025ci_cd_android} and Enemosah et al.~\cite{enemosah2025enhancing} examined CI practices in Android projects, reporting inconsistent pipeline structures and low automation rates for deployment ($\approx$9\%). Similar gaps in practice were observed in Travis CI configurations, where most workflows focused on build/test while neglecting deployment~\cite{gallaba2020tse}. Improper CI setups have been linked to longer build times~\cite{ghaleb2019empirical,ghaleb2022studying} and unsuccessful builds~\cite{ghaleb2019studying,ghaleb2022studying}.

Empirical studies have also highlighted workflow maintenance challenges. Valenzuela-Toledo et al.~\cite{valenzuela2024hidden} found that GHA workflows are frequently updated, even though they are relatively short ($\approx$60 lines), with changes often targeting bugs or CI improvements. Other research identified workflow smells, such as hard-coded versions, excessive permissions, and redundant logic~\cite{khatami2024catching,zampetti2020empirical,bouzenia2024resource}, which signal suboptimal practices and hinder maintainability.

In the context of ML projects, studies have found poor adoption of standardized testing and bad practices such as hard-coded dependencies, leading to brittle pipelines~\cite{rzig2024empirical}. Faqih et al.~\cite{faqih2024empirical} classified GHA tools and showed that build automation is the most prevalent category, while test automation is the least common. Golzadeh et al.~\cite{golzadeh2022rise} confirmed GHA's growing dominance on GitHub.

Finally, multiple studies call for clearer guidelines and tooling to support CI workflow creation and evolution~\cite{valenzuela2024hidden,khatami2024catching}. Despite growing interest, prior work has not jointly analyzed workflow complexity, heterogeneity, and compliance with CI principles at scale. Our study addressed this gap by offering a thorough analysis of GHA workflows across diverse projects.

\section{Study Setup}
\label{sec:Study_Setup}

\subsection{Dataset}
\label{data_collection}
    We utilized the official GitHub REST API to identify public repositories with Java, Python, or C++ as their dominant language that contain CI/CD configuration files. To ensure that we include mature and well-maintained repositories, we have followed established criteria to filter out inactive and toy projects~\cite{beller2017travistorrent}, by selecting non-archive, non-fork projects with $\ge$ 10 watchers (stars). We further restricted our selection to repositories created on or before December 31, 2024, to avoid including newly created or short-lived projects. After that, we selected only the repositories that adopt GHA. To do that, based on the documentation, we identified the GHA YAML-based workflow files that are located in the standard \texttt{.github/workflows} directory. Then, we filtered out repositories that have fewer than 50 GHA build runs to ensure that they include consistently active projects~\cite{beller2017travistorrent}. We developed Python scripts to automate the entire data collection and filtration process. This produced a dataset of repositories that represent actively maintained, non-trivial projects with meaningful CI activity. Table~\ref{tab:ci_summary} gives some basic statistics of the number of resulting repositories and YAML files.

    \begin{table}[ht]
        \centering
        \caption{Overview of our dataset}
        \label{tab:ci_summary}
            \begin{tabular}{lrrr}
                \toprule
                \textbf{Language} & \textbf{\# Filtered Projects} & \textbf{\# YAML Files} & \textbf{Avg. YAMLs/Project} \\
                \midrule
                Java   & 1,576 & 5,749  & 3.6 \\
                Python & 4,428 & 15,394 & 3.5 \\
                C++    & 1,664 & 6,720  & 4.0 \\
                \midrule
                \textbf{Total} & \textbf{7,668} & \textbf{27,863} & \textbf{3.6} \\
                \bottomrule
            \end{tabular}%
    \end{table}

\section{Empirical Evaluation}
\label{sec:Evaluation}

\begin{figure*}
    \centering
    \vspace{-5pt}
    \includegraphics[width=1\textwidth]{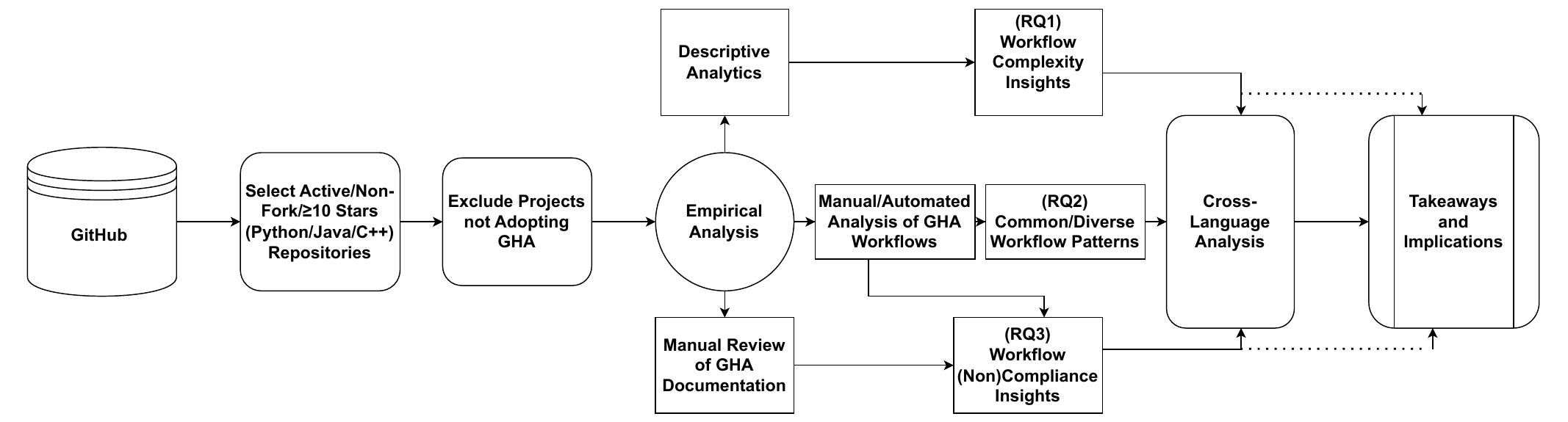} 
    \caption{Overview of our study}
    \vspace{-5pt}
    \label{fig:study_overview} 
\end{figure*}

Figure~\ref{fig:study_overview} presents an overview of our study design.

We conducted a quantitative comparative analysis through static inspection of GHA workflow YAML files collected from open-source Java, Python, and C++ projects, as summarized in Table~\ref{tab:ci_summary}. 

Our methodology combines descriptive statistical analysis to characterize workflow complexity. Manual inspection was performed to establish a statistically representative sample of YAML configurations, through which we identified recurring structural patterns and job-step sequences. 

Workflows exhibiting similar patterns were grouped to systematically examine structural characteristics, configuration heterogeneity, and compliance with established CI best practices.

\subsection{\textbf{RQ1: How complex are GHA workflows in open-source projects?}}

\subsubsection{Motivation}
CI pipelines have become the operational backbone of modern open-source development, yet their internal complexity is rarely measured or compared in a systematic way. A GHA workflow file is not just a configuration artifact, it is an executable infrastructure that every contributor must be able to read, extend, and debug. Poor CI practices measurably increase build failure frequency and delay their resolution~\cite{zampetti2020empirical}. When a workflow grows too large, nests too deeply, or calls on dozens of third-party actions, it becomes harder to maintain, more prone to silent failures, and less accessible to newcomers, slowing onboarding and creating bottlenecks in resolving issues. However, it remains unclear whether the primary programming language of a project shapes how developers design their automation pipelines, a question with direct implications for tooling, documentation, and community support strategies~\cite{valenzuela2025explaining}.  Measuring workflow complexity across Java, Python, and C++ projects gives us the empirical evidence needed to answer that question and to expose the structural patterns that tool designers and DevOps practitioners should account for.

\subsubsection{Approach}

\smallskip\noindent\textbf{Complexity Metrics.}
We analyzed GHA workflow files drawn from open-source repositories spanning three programming languages: Java, Python, and C++. Complexity was quantified along four complementary dimensions (Table~\ref{tab:variables}):

\begin{itemize}
  \item \textbf{Workflow scale metrics}: lines of YAML per workflow file, number of jobs, and total number of steps, which together capture overall scale of workflow configuration.
  \item \textbf{Structural-depth metrics}: maximum YAML nesting depth, maximum sequential steps within any single job (the longest chain of operations a runner must complete without branching), workflow dependency depth (the longest job dependency chain), and job parallelism (how many jobs can execute simultaneously).
  \item \textbf{Advanced workflow features}: binary and count indicators for matrix build strategies, conditional execution (\texttt{if:} clauses at job and step level), job-dependency declarations (\texttt{needs:}), reusable workflow calls, local custom actions, and external Marketplace actions.
  \item \textbf{Derived complexity ratios}: average steps per job, dependency ratio, external-action diversity (unique external actions relative to total steps), and conditional density, which normalize raw counts to enable fair cross-language comparison.
\end{itemize}

\smallskip\noindent\textbf{Complexity Classification.}
All metrics were extracted automatically using a custom Python script applied to each parsed YAML file. Workflows that could not be parsed due to syntax errors, or that lacked a \texttt{jobs:} block, the required section defining executable pipeline logic in \textit{GHA}, were excluded. Such files do not represent functional CI pipelines and would otherwise bias structural metrics toward zero.
Workflow complexity levels were classified using value-based threshold bands derived from the full dataset distribution of each metric: values at or below the 70th percentile were labeled \emph{low}, values between the 70th and 80th percentiles \emph{moderate}, values between the 80th and 90th percentiles \emph{high}, and values above the 90th percentile \emph{very high}. Using value thresholds rather than rank-based assignment ensures that identical metric values consistently fall within the same band~\cite{alves2010deriving}.

\smallskip\noindent\textbf{Statistical Analysis.}
Cross-language differences were evaluated using the Kruskal-Wallis $H$-test~\cite{kruskal1952use} for overall comparisons, followed by pairwise Mann-Whitney $U$ tests~\cite{mann1947test}. Cliff's $\delta$ was used as the standardized effect size~\cite{romano2006appropriate,grissom2005effect}, with statistical significance determined at $p < 0.05$~\cite{ghaleb2019empirical}.

\begin{longtable}{>{\raggedright\arraybackslash}p{2.5cm}|
                  >{\raggedright\arraybackslash}p{4.6cm}|
                  >{\raggedright\arraybackslash}p{3.6cm}}
\caption{Variable Definitions \label{tab:variables}}\\

\multicolumn{3}{l}{\textit{\textbf{Workflow Scale Metrics}}} \\[2pt]
\toprule
\textbf{Metric} & \textbf{Explanation} & \textbf{Importance} \\
\midrule
\endfirsthead

\multicolumn{3}{c}{{\tablename\ \thetable{} (continued)}}\\
\toprule
\textbf{Metric} & \textbf{Explanation} & \textbf{Importance} \\
\midrule
\endhead

\bottomrule \multicolumn{3}{r}{{Continued on next page}} \\
\endfoot

\bottomrule
\endlastfoot

Lines Of Yaml & Total lines in the workflow YAML file. & Workflow size proxy for comparing complexity. \\
Matrix Size & Number of job variants from matrix builds. & Bigger matrices = more parallel test complexity. \\
Num Jobs & Total jobs in the workflow. & More jobs = greater pipeline complexity. \\
Num Steps & Total steps across all jobs. & More steps = larger workload complexity. \\

\midrule
\multicolumn{3}{l}{\textit{\textbf{Structural Depth Metrics}}} \\[2pt]
\midrule
Job Parallelism & Max jobs running concurrently. & Higher parallelism = more resource complexity. \\
Max Nesting Depth & Max YAML nesting level. & Deeper nesting = harder to understand structure. \\
Max Sequential Steps & Max steps within a job. & Largest job highlights bottlenecks. \\
Num Job Dependencies & Total job dependencies (needs). & More dependencies = more structured pipeline. \\
Vertical Depth & Longest dependency chain. & Longer dependency chains = more sequential complexity. \\

\midrule
\multicolumn{3}{l}{\textit{\textbf{Advanced Workflow Features}}} \\[2pt]
\midrule
Has Conditionals & Presence of conditional (if:) statements. & Conditionals add execution complexity. \\
Has Job Dependencies & Presence of job dependencies. & Job dependencies add structural complexity. \\
Has Matrix & Uses build matrix. & Matrix builds multiply workflow complexity. \\
Num Conditionals & Count of conditional statements. & More conditionals = higher branching complexity. \\
Num Local Actions & Count of local actions. & More custom actions = higher internal maintenance. \\
Num Marketplace Actions & Count of marketplace actions. & More marketplace actions = higher integration complexity. \\
Num Reusable Workflows & Times invoked by other workflows. & Reuse reduces complexity via modularization. \\
Num Unique External Actions & Unique external actions used. & More diverse actions = broader external dependencies. \\
Uses External Actions & Uses any external actions. & External actions increase maintenance complexity. \\
Uses Local Actions & Uses local actions. & Local actions increase development complexity. \\
Uses Reusable Workflows & Uses reusable workflows. & Reusable workflows reduce per-workflow complexity. \\

\midrule
\multicolumn{3}{l}{\textit{\textbf{Derived Complexity Ratios}}} \\[2pt]
\midrule
Avg Steps Per Job & Avg steps per job. & Higher steps/job = more complex individual jobs. \\
Conditional Density & Conditionals per job. & More conditionals/job = denser branching logic. \\
Dependency Ratio & Dependencies per job. & Higher coupling = more sequential workflow structure. \\
External Action Diversity & Unique/total external actions ratio. & More external tools = higher dependency complexity. \\

\end{longtable}

\subsubsection{Findings}

\begin{figure}[!t]
  \centering
  \includegraphics[width=\columnwidth]{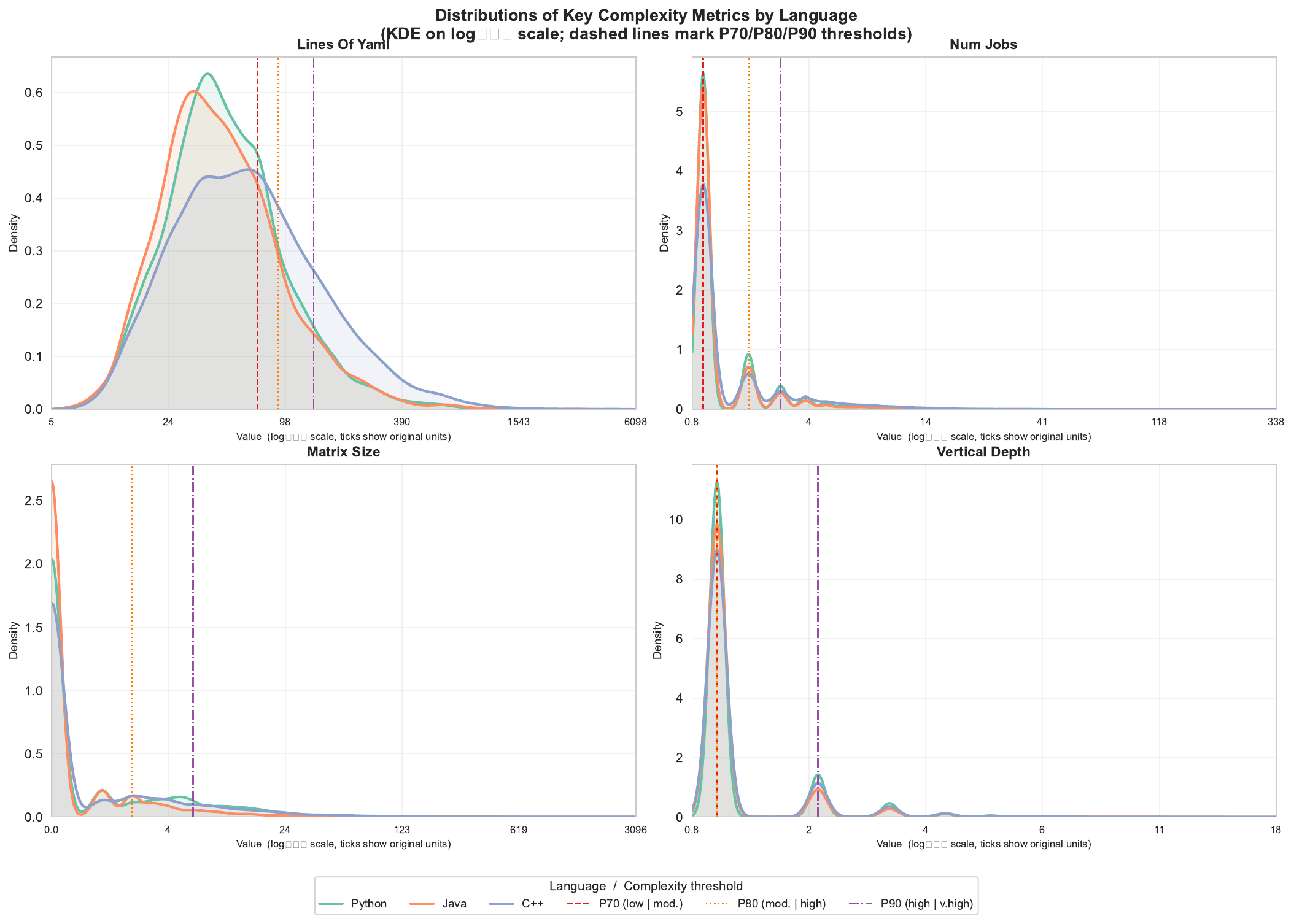}
  \caption{Kernel density estimates (KDE) of four core size metrics, lines of YAML, number of jobs, matrix size, and vertical depth, for Python (green), Java (orange), and C++ (blue), plotted on a $\log_{1+x}$ scale with back-transformed tick labels. Vertical lines mark the P70, P80, and P90 complexity-band thresholds.}
  \label{fig:dist}
\end{figure}

\smallskip\noindent\textbf{Workflow scale metrics.} Figure~\ref{fig:dist} presents kernel density estimates (KDE) of four foundational size metrics across all the three studied languages. The distributions are plotted on a $\log_{1+x}$ scale to mitigate the influence of extreme outliers that would otherwise compress the visible
range of the data. To maintain interpretability, axis tick labels are back transformed to their original values. Three vertical reference lines indicate the percentile based complexity thresholds applied in this study: P70 (red dashed),P80 (orange dotted), and P90 (purple dash-dot).

In every panel the density curves are sharply right-skewed: each language exhibits an early peak followed by a long right tail, confirming that most workflows are lean while a small minority are substantially more complex. For lines of YAML, the three language curves peak between roughly 30 and 90 lines and the P70 reference line falls at 70 lines, meaning 70\,\% of all workflows sit to the left of that boundary regardless of language. The P90 line at 137 lines marks the onset of the \emph{very high} band; only 10\,\% of the total data extends beyond it, yet the density tails continue well to the right, reaching extremes of 4,517 lines (Python), 2,975 lines (Java), and 2,869 lines (C++). The C++ curve (blue) peaks at a noticeably higher value and decays more slowly than Java (orange) and Python (green), visually confirming that C++ workflows are systematically larger across the bulk of the distribution.

The number of jobs panel tells a stark story: all three density curves spike sharply at a log-transformed value near zero (corresponding to 1 job) and become effectively flat beyond that point. The P70 reference line sits at 1 job, meaning the majority of all workflows run exactly one job with no parallelism. The rare workflows with tens or hundreds of jobs are visible as faint density far to the right but represent only a small fraction of the dataset. The vertical depth panel mirrors this pattern almost exactly: a sharp peak at depth 1 with the P70 and P80 boundaries both coinciding at 1, confirming that most pipelines consist of fully independent jobs with no sequencing logic between them. 

Finally, the matrix size panel shows the same concentration near zero, with the P70 line itself at 0, indicating that the majority of workflows do not use matrix builds at all; the visible density observed above zero corresponds entirely to the minority of workflows that utilize matrix strategies for multi-environment testing.

C++ workflows are the largest of the three languages. Their median
of 58 lines is 38\,\% greater than the Java median of 42 lines and 29\,\%
greater than the Python median of 45 lines, a gap that persists after
removing the influence of extreme outliers that inflate the mean.
The upper tail reinforces this: the C++ 90th-percentile threshold reaches 196 lines, well above the overall dataset threshold of 137 lines, whereas Java (115 lines) and Python (120 lines) both sit below it. Step counts follow the same ordering: C++ workflows have a median of 6 steps, 50\,\% more than the Java median of 4 and 20\,\% more than Python's median of 5.  Both gaps carry small but consistent effect sizes, as detailed in Table~\ref{tab:stats}.

\begin{figure*}[!t]
  \centering
  \includegraphics[width=\textwidth]{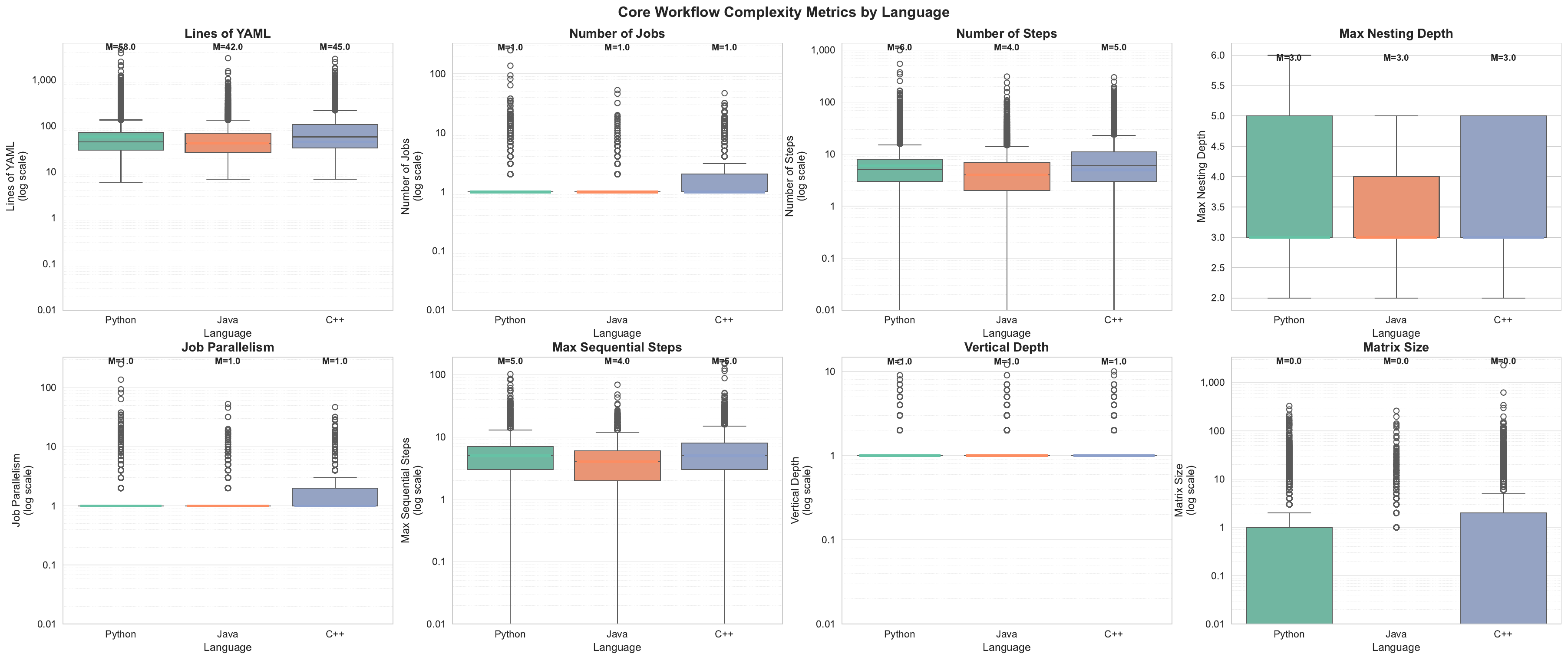}
  \caption{Box plots of all eight core workflow complexity metrics by language, with per-language medians ($M$) annotated.}
  \label{fig:core}
\end{figure*}

Figure~\ref{fig:core} presents box plots for all eight core metrics
simultaneously. The log-scale y-axis and bold median lines make the ordering immediately legible: C++ boxes sit higher than Java and Python for
lines-of-YAML ($M = 58$ vs $42$, $45$), number of steps ($M = 6$ vs $4$, $5$), and max-sequential-steps ($M = 5$ vs $4$, $5$), while all three languages share identical medians for number of jobs, job-parallelism, and vertical depth ($M = 1$ in every case). The dense outlier clouds above every box are consistent with the long right tails visible in the KDE panels of Fig.~\ref{fig:dist}, confirming the heavy tailed character of every complexity metric regardless of language.

\smallskip\noindent\textbf{Workflow Simplicity.} Most workflows are small and structurally flat.
This pattern reflects the dominant role of CI in open-source development: automating discrete, self-contained tasks such as running tests, building packages, or checking code style, rather than orchestrating end-to-end delivery sequences. Small, flat workflows are easy to read, straightforward to debug, and accessible to new contributors. The downside is that jobs which could safely run in parallel are instead executed sequentially, wasting available speedup. For small or moderately active projects this trade-off is reasonable; for larger, high-frequency repositories, the cumulative cost of underutilized parallelism becomes a real bottleneck.

\smallskip\noindent\textbf{Language Size Differences.} C++ workflows are measurably larger and step-heavier than those of Java or Python.
These differences are consistent across the other five correlated metrics reported in Table~\ref{tab:stats}.

\begin{figure}[ht]
  \centering
  \includegraphics[width=\columnwidth]{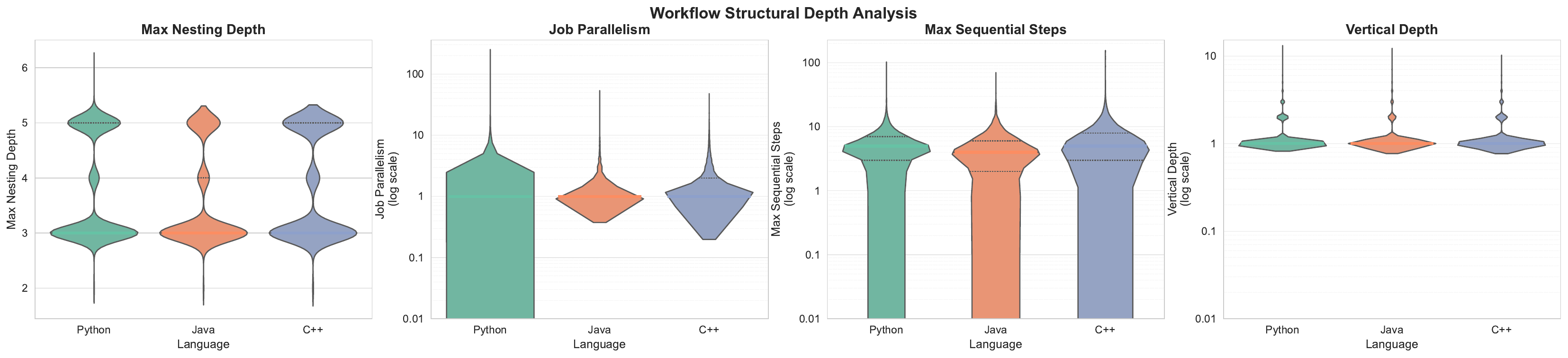}
  \caption{Violin plots for the four structural-depth metrics: max-nesting-depth, job-parallelism, max-sequential-steps, and vertical-depth. Internal dashed lines mark the quartiles within each violin.}
  \label{fig:depth}
\end{figure}

\smallskip\noindent\textbf{Structural Depth.}
Figure~\ref{fig:depth} uses violin plots to reveal the full shape of the four structural-depth distributions. The violin for max-nesting-depth is virtually identical across all three languages: a wide, symmetric mass centered at 3 to 5 levels with a hard upper boundary at 6. Practically all workflows (99.9\,\%) fall within the \emph{low} band for this metric.  While the Kruskal-Wallis test detects a statistically significant overall difference ($H = 216.5$, $p < 10^{-47}$), all pairwise Cliff's $\delta$ values remain below 0.13 indicating negligible effect sizes. This suggests that while small differences are detectable at the scale of the dataset, they are not practically meaningful.
The job-parallelism and vertical-depth violins are nearly degenerate, with
large, dense masses concentrated at value 1 and very thin tails extending upward. Approximately 85\% of all workflows fall within the \emph{low} band for vertical depth, indicating that most jobs have no dependency on one another at all. Consequently, the majority of workflows operate as flat collections of tasks rather than structured, multi-stage pipelines. Only 4.8\,\% of workflows overall, and at most 5.3\,\% within any single language, exhibit the deeper dependency structures typical of fully orchestrated delivery pipelines.
The max-sequential-steps panel is the structural standout.  The C++ violin body is visibly wider and taller, with a median of 5 versus 4 for Java and Python, and a 90th-percentile threshold of 12 steps versus 9 for the other two languages. The pairwise C++ vs Java comparison yields a \emph{small} effect size ($\delta = 0.186$, $p < 10^{-81}$), reflecting the multi-stage build, link, and test ceremony that native compiled projects inherently require.

\smallskip\noindent\textbf{Orchestration Depth.} Job orchestration is shallow across the dataset, and C++ jobs run the longest internal step chains.
Only 4.8\,\% of workflows overall, and at most 5.3\,\% within any single language, exhibit the deeper dependency structures typical of fully orchestrated delivery pipelines, reinforcing the dataset's overall flatness. Where sequential depth does appear, it concentrates in C++, consistent with the multi-stage build, link, and test sequence native compilation requires.

\begin{figure}[!t]
  \centering
  \includegraphics[width=\columnwidth]{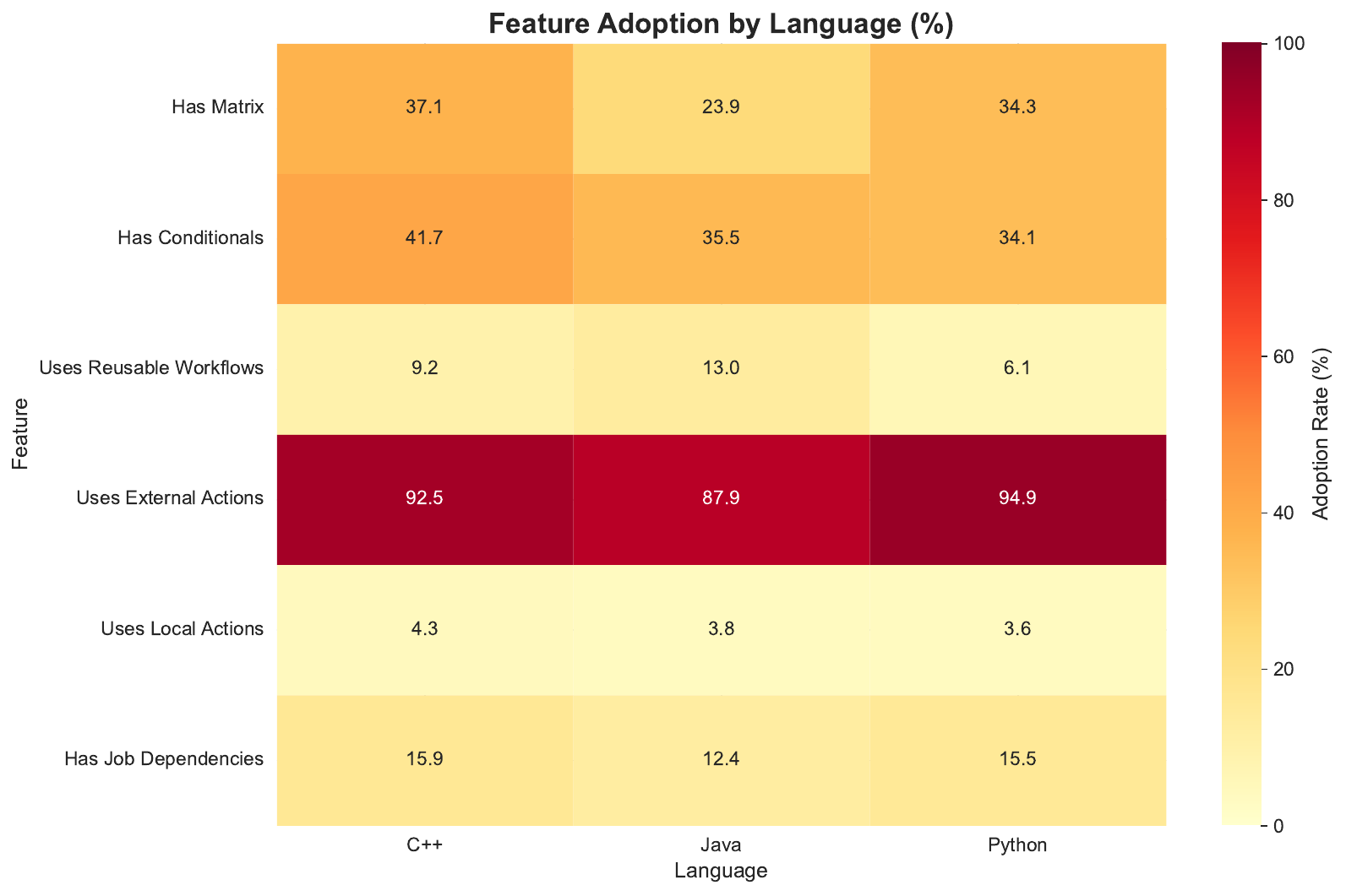}
  \caption{Heatmap of binary feature adoption rates (\%) across the three languages. Cell color encodes adoption intensity on a 0\% to 100\,\% scale (pale yellow = rare, deep red = near-universal).}
  \label{fig:features}
\end{figure}

\smallskip\noindent\textbf{Advanced Workflow features.}
Figure~\ref{fig:features} presents the feature adoption heatmap, where color intensity immediately signals which features dominate: a single deep-red row stands out from the rest, which stay firmly in pale yellow and light orange.

\textit{External actions} dominate universally: 94.9\,\% of Python, 92.5\,\% of C++, and 87.9\,\% of Java workflows use at least one externally published action, an overall rate of 92.9\,\%.  This near-universal take-up establishes the GitHub Marketplace as the primary mechanism through which automation is composed in open-source GHA workflows. It also introduces a pervasive supply-chain dependency: the overwhelming majority of these workflows trust and execute code from external maintainers at every CI run.
This makes the GitHub Marketplace the de-facto standard library of open-source automation, reflecting how effectively the Marketplace model has solved the discoverability and reuse problem for common CI tasks.

\textit{Matrix builds} exhibit clear language-specific adoption. C++ leads at 37.1\,\%, followed by Python (34.3\,\%), while Java trails at 23.9\,\%, a 13-percentage-point gap. This reflects differences in ecosystem fragmentation: C++ must account for compiler, operating system, and standard library variations, while Python projects commonly test across multiple interpreter versions. Java's JVM largely abstracts platform differences, reducing the need for extensive matrix testing. Although matrix builds improve reliability by detecting environment-specific defects, they also increase CI cost (e.g., a $3 \times 3$ matrix requires nine times more runner executions), highlighting the need for cost-aware matrix governance and guidance on selecting appropriate test configurations.

\textit{Conditional execution} (\texttt{if:} clauses) is used by 41.7\,\% of C++ workflows, above Java (35.5\,\%) and Python (34.1\,\%). The mean
conditional count is also highest in C++ (1.78 per workflow) versus Java (0.99) and Python (0.95), with 13.7\,\% of C++ workflows in the \emph{very high} band. Conditional logic is especially valued in C++ where skipping expensive compilation steps has a tangible time-saving effect.

\textit{Reusable workflows} remain a minority feature across all ecosystems. Java shows the highest adoption (13.0\,\%), followed by C++ (9.2\,\%) and Python (6.1\,\%), potentially reflecting the greater prevalence of monorepos in Java projects. Similarly, local custom actions ($\leq 4.3\%$) and deep job-dependency chains ($\leq 5.3\%$ in the \emph{very high} band) are rare. While basic action composition is common, advanced modularity and reuse mechanisms remain underutilized, with most workflows implemented as standalone pipelines rather than shared, reusable components. This highlights a gap between GHA's modularity capabilities and current open-source practice, providing a baseline for tracking future adoption.

\smallskip\noindent\textbf{Feature Adoption Pattern.} Adoption of GHA's advanced features is bimodal: low-effort features are near-universal, architecturally demanding ones are not.
External actions require no design decision beyond picking one from the Marketplace, and adoption is accordingly near-universal. Matrix strategies, conditional logic, and reusable workflows each require the developer to design something (a test matrix, a branching condition, a shared interface), and adoption of each drops sharply and varies by language. The dividing line is not feature complexity but whether the feature demands upfront design investment from the workflow author.

\begin{figure}[ht]
  \centering
  \includegraphics[width=\columnwidth]{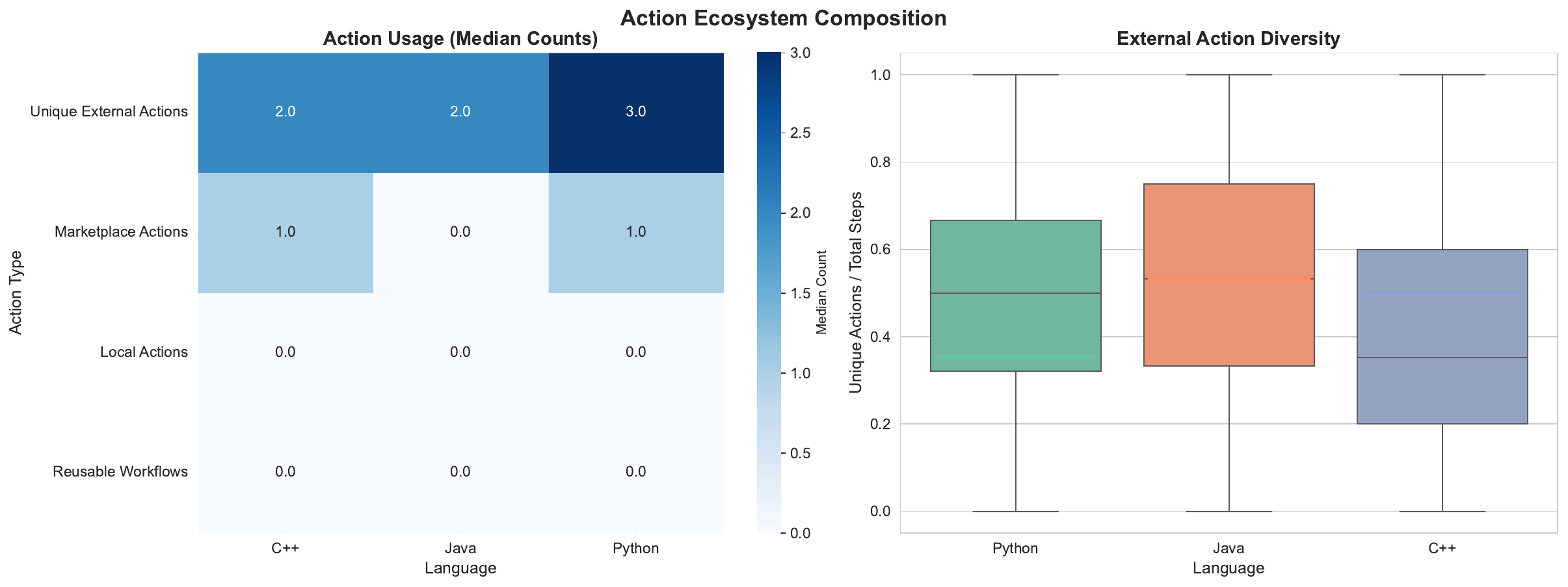}
  \caption{Left: heatmap of median action counts by category and language. Right: box plot of external-action diversity (unique external actions per total step) by language.}
  \label{fig:actions}
\end{figure}

\smallskip\noindent\textbf{Action Ecosystem Composition.}
Figure~\ref{fig:actions} shows action usage by language (left) and external-action diversity (right). Marketplace and unique external actions are the only action types with meaningful median counts, while local actions and reusable workflows have median usage of zero across all languages. Python workflows use the most unique external actions (median = 3), compared with C++ and Java (median = 2), with a 90th-percentile value of 5 across all ecosystems.
Despite similar action counts, Java exhibits the highest external-action diversity (median = 0.53), followed by Python (0.50) and C++ (0.35). This metric shows the strongest cross-language difference of all 18 complexity metrics ($H = 765.7$, $p < 10^{-166}$), with small but significant effects between C++ and Java ($\delta = -0.234$) and C++ and Python ($\delta = -0.212$). The pattern suggests that Java and Python rely more heavily on specialized Marketplace actions, whereas C++ workflows perform more tasks through shell-script steps.

\smallskip\noindent\textbf{Action Diversity.} Java and Python workflows use external actions more diversely than C++ workflows.
Java and Python developers reach for a different purpose-built action at each step because well supported Marketplace actions exist for runtime setup, dependency caching, testing, and package publishing in both ecosystems~\cite{golzadeh2022rise}. C++ has no comparable breadth of language specific Marketplace coverage,
leaving developers to implement equivalent logic in raw \texttt{run:} blocks, producing workflows where many consecutive steps are shell commands surrounding only a small number of action calls. The diversity gap is therefore a structural consequence of ecosystem maturity, not a
difference in developer intent, and it points to where targeted Marketplace investment would have the greatest impact on C++ pipeline quality.

\begin{figure}[ht]
  \centering
  \includegraphics[width=\columnwidth]{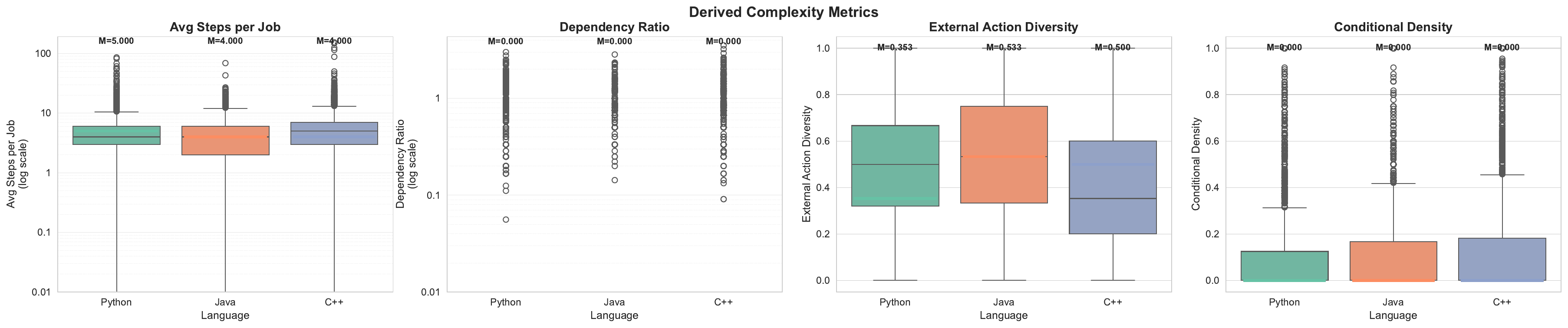}
  \caption{Box plots for the four derived, size-independent complexity metrics by language. Median annotations ($M$) appear at the top of each panel.}
  \label{fig:derived}
\end{figure}

\smallskip\noindent\textbf{Derived Complexity Ratios.}
Figure~\ref{fig:derived} presents four size-independent complexity metrics. Dependency ratio and conditional density have median values of zero across all languages, indicating that most workflows remain dependency-free and rarely use conditional execution, although a small number of outliers exhibit highly complex dependency chains and branching logic.
Average steps per job mirrors the raw step-count results, with C++ showing the highest median (5) compared with Java and Python (4), confirming that C++ workflows are more step-intensive on a per-job basis ($\delta = 0.161$ vs. Java). External-action diversity follows the same pattern observed earlier: Java (median = 0.53) and Python (0.50) exceed C++ (0.35), suggesting greater reliance on specialized Marketplace actions, while C++ implements more functionality through shell-script steps.

\smallskip\noindent\textbf{Normalized Ratios.} Size-independent ratios confirm that C++'s larger footprint and Java/Python's action diversity are structural, not artifacts of workflow size.
Normalizing for size does not erase the earlier differences; it sharpens them. C++ remains more step-intensive per job and Java and Python remain more action-diverse per step even after removing raw scale as a confound, indicating genuine compositional differences between ecosystems rather than an artifact of C++ workflows simply being longer.

\smallskip\noindent\textbf{Statistical Significance and Effect Sizes.}
Kruskal-Wallis testing detected statistically significant cross-language
differences for all 18 metrics ($p < 0.05$ in every case; $p < 10^{-8}$ for
16 of 18), demonstrating that language has a reliable,
measurable association with workflow complexity.
Table~\ref{tab:stats} summarizes pairwise results for the seven comparisons
that crossed into \emph{small} effect-size territory.

Despite pervasive statistical significance, the majority of pairwise
comparisons yield only \emph{negligible} Cliff's $\delta$ values
($|\delta| < 0.147$), illustrating the important distinction between
statistical and practical significance in large-scale empirical studies.
The small-effect comparisons form a coherent cluster: all involve C++ versus Java or Python, all point in the same direction (C++ larger or more step-heavy; Java and Python higher in action diversity), providing strong evidence of a systematic ecosystem-rooted difference rather than random noise.

\smallskip\noindent\textbf{Significance vs.\ Practical Relevance.} Statistical significance is pervasive at this dataset scale, but only a handful of comparisons are practically meaningful.
At $N=27{,}863$, even small language differences reach statistical significance; effect size, not the $p$-value, is therefore the more informative signal in this study. Every small-or-larger effect points in the same direction across metrics rather than scattering unpredictably, evidence that language captures a real structural distinction rather than sampling noise.

\begin{table}[t]
  \caption{Pairwise Mann-Whitney $U$ test results for metrics with small or
           greater effect sizes. Cliff's $\delta$: $|\delta| < 0.147$ = negligible; $0.147$ to $0.33$ = small. All other pairwise comparisons are statistically significant ($p < 0.05$) but yield negligible effects.}
  \label{tab:stats}
  \centering
  \small
  \setlength{\tabcolsep}{4pt}
  \begin{tabular}{llccccc}
    \toprule
    \textbf{Metric} & \textbf{Comparison}
      & \textbf{Med\textsubscript{1}}
      & \textbf{Med\textsubscript{2}}
      & \textbf{Cliff's $\delta$}
      & \textbf{Effect}
      & \textbf{$p$-value} \\
    \midrule
    Lines of YAML        & C++ vs Java   & 58   & 42   & $+0.217$ & small & $< 0.0001$ \\
    Lines of YAML        & C++ vs Python & 58   & 45   & $+0.172$ & small & $< 0.0001$ \\
    Num Steps            & C++ vs Java   &  6   &  4   & $+0.205$ & small & $< 0.0001$ \\
    Max Sequential Steps & C++ vs Java   &  5   &  4   & $+0.186$ & small & $< 0.0001$ \\
    Avg Steps / Job      & C++ vs Java   &  5   &  4   & $+0.161$ & small & $< 0.0001$ \\
    Action Diversity     & C++ vs Java   & 0.35 & 0.53 & $-0.234$ & small & $< 0.0001$ \\
    Action Diversity     & C++ vs Python & 0.35 & 0.50 & $-0.212$ & small & $< 0.0001$ \\
    \bottomrule
  \end{tabular}
\end{table}
    
\vspace{-6pt}
\begin{rqbox}
\textbf{RQ\textsubscript{1} Summary.}
GHA workflows in open-source projects are typically small, flat, single-job pipelines built around external Marketplace actions. C++ workflows are consistently larger, more step-heavy, and rely more on matrix builds than Java or Python, while Java and Python compose more diverse action ecosystems. Advanced orchestration features, reusable workflows, local actions, and deep job dependencies, remain largely unused across all three languages.
\end{rqbox}

\subsection{\textbf{RQ2: What are the common and diverse structures and usage patterns in GHA workflows?}}

\subsubsection{Motivation}
CI configurations in open-source projects are often built independently, with little coordination, resulting in inconsistent and non-standardized pipelines. While prior work has documented wide variation in CI practices across projects, it remains largely unclear which structural patterns are commonly adopted, which are inconsistent, and where standardization gaps persist. These inconsistencies carry real costs, such as misconfigured pipelines, automation failures, and wasted debugging time~\cite{chandrasekara2021getting}. Understanding these structural patterns is therefore essential for developers and tool designers aiming to improve reliability and reduce inefficiencies. In this RQ, we systematically analyze GitHub Actions workflows to assess their degree of structural and semantic standardization, evaluating their conformance to recurring patterns and identifying gaps that signal opportunities for reusable templates and shared best practices across open-source CI ecosystems.

\subsubsection{Approach}

\smallskip\noindent\textbf{Workflow Analysis.}
We analyzed 382 GHA workflows spanning three programming languages: Python (\(n=222\), 58.1\%), C++ (\(n=84\), 22.0\%), and Java (\(n=76\), 19.9\%). Our analysis focuses on identifying recurring step constructs~(e.g., checkout, setup, test), common step sequences~(e.g., checkout → setup → install → test), structural deviations, and the adoption of advanced GHA features. To characterize CI usage, we examine both the prevalence of these constructs and the extent to which workflows adopt modularity mechanisms, such as reusable workflows and composite actions. For features amenable to static analysis, including version pinning, caching, setup actions, trigger usage, and workflow modularity, we employ a standard YAML parser to automatically extract and categorize patterns.

\smallskip\noindent\textbf{Manual Coding and Classification.}
To uncover structural and naming patterns not directly detectable through static analysis, we employ open coding~\cite{corbin1990grounded}. The two authors independently manually analyzed a statistically significant stratified random sample of 382 workflows (with 95\% confidence level and $\pm$5 margin of error). Since step names vary widely across projects, we develop semantic labels that capture the intent of each step regardless of how it is named. For instance, any step whose purpose is to check out code is grouped under the label \textit{checkout}, and similarly for \textit{install}, \textit{build}, and others.
To ensure consistency, the two authors independently annotated the sample and then compared their semantic label assignments. Initial agreement stood at 85\%, yielding a Cohen's Kappa of $\kappa = 0.82$, indicating strong agreement~\cite{landis1977}. Disagreements were resolved through structured discussion until consensus was reached. The finalized label set was then encoded as regular expression patterns and action slug mappings in the classification script and applied to the full dataset. This iterative process was repeated until all sampled YAML files were consistently labeled, after which the classifier was validated against a held-out subset to confirm agreement with the manual annotations. Based on this, we define rules (regular expressions and keywords) to automatically label components across the full dataset. Patterns are considered common if they appear in at least 5\% of the dataset. Deviations from typical step sequences identified in the GHA documentation are further analyzed in context to examine whether they are associated with specific project types, languages, or usage patterns.

\smallskip\noindent\textbf{Statistical Analysis.}
Cross-language differences in continuous structural metrics (sequence length, step count, completeness score) were evaluated with the Kruskal-Wallis $H$-test, followed by pairwise Mann-Whitney $U$ tests with Cliff's $\delta$ effect sizes~\cite{kruskal1952use,mann1947test,romano2006appropriate}. For binary adoption features (e.g., build, test, cache, and SHA pinning), we used chi-square tests~\cite{pearson1900criterion} for overall language comparisons and pairwise Fisher's exact tests~\cite{fisher1922interpretation}; Cohen's $h$~\cite{cohen1988statistical} served as the effect size for proportions, with $|h| < 0.2$ interpreted as negligible. Significance was assessed at $p < 0.05$, consistent with RQ1.

\subsubsection{Findings}

\smallskip\noindent\textbf{Frequent Step Category}.
We identified \textbf{21 semantic step categories} that appear in
at least 5\% of workflows (Table~\ref{tab:category}).
\textit{Checkout Repository} is the most prevalent semantic step category, present in 91.6\% of workflows, confirming its role as a
near universal entry point for CI pipelines and a foundational prerequisite for subsequent steps.
\textit{Setup Runtime Environment} (58.6\%) and
\textit{Install Project Dependencies} (45.3\%) form the next tier,
reflecting the standard pattern of environment preparation before
compilation or testing.
\textit{Execute Tests} appears in only 27.8\% of
workflows, substantially lower than \textit{Build Project}
(44.2\%), indicating that many pipelines prioritize build
verification and artifact production over automated testing.
Security oriented category \textit{Perform Security Analysis}
(9.7\%) and \textit{Initialize Security Analysis}
(8.6\%) together appear in roughly one in ten workflows, pointing
to growing but still minority adoption of automated security
scanning in CI.
 
\begin{table}[t]
  \centering
  \caption{Semantic Step appearing in $\geq$5\% of workflows
           (\(N=382\)).}
  \label{tab:category}
  \small
  \begin{tabular}{lrr}
    \toprule
    \textbf{Semantic Name} & \textbf{Count} & \textbf{\%} \\
    \midrule
    Checkout Repository           & 350 & 91.6 \\
    Setup Runtime Environment     & 224 & 58.6 \\
    Install Project Dependencies  & 173 & 45.3 \\
    Build Project                 & 169 & 44.2 \\
    Execute Tests                 & 106 & 27.8 \\
    Upload Build Artifacts        &  98 & 25.6 \\
    Run Static Analysis           &  84 & 22.0 \\
    Setup Build Environment       &  60 & 15.7 \\
    Manage Build Cache            &  46 & 12.0 \\
    Configure Build Profile       &  44 & 11.5 \\
    Inspect Runtime Environment   &  39 & 10.2 \\
    Perform Security Analysis     &  37 &  9.7 \\
    Publish Package               &  33 &  8.6 \\
    Initialize Security Analysis  &  33 &  8.6 \\
    Download Build Artifacts      &  32 &  8.4 \\
    Publish Release               &  30 &  7.8 \\
    Build Source Code             &  28 &  7.3 \\
    Generate Version Metadata     &  25 &  6.5 \\
    Generate Test Coverage        &  25 &  6.5 \\
    Upload Coverage               &  23 &  6.0 \\
    Build Package                 &  23 &  6.0 \\
    \bottomrule
  \end{tabular}
\end{table}
 
\smallskip\noindent\textbf{Sequential Structure of CI Workflows}.
The average workflow sequence contains \textbf{5.4 steps}
(median~=~5.0), suggesting pipelines are intentionally compact.
Only \textbf{one full sequence} meets the global 5\% threshold:
 
\begin{quote}
\textit{Checkout Repository $\rightarrow$ Initialize Security
Analysis $\rightarrow$ Build Project $\rightarrow$ Perform Security
Analysis} \quad (22 workflows, 5.8\%)
\end{quote}
 
\noindent This sequence reflects the CodeQL scanning
pattern~\cite{codeql_overview}, where security analysis
steps bracket the build step in the workflow and is observed predominantly in Python (16 occurrences, 7.21\%) and Java (5 occurrences, 6.58\%).
The rarity of globally repeating full sequences confirms that exact
pipeline configurations are highly project specific; category level
frequency is therefore the more appropriate unit for cross-project
pattern analysis~\cite{decan2022use,hilton2016usage}.

\smallskip\noindent\textbf{Reproducible Pattern.} The only globally reproducible workflow pattern is a security analysis pipeline, not a build-and-test pipeline.
Its structural consistency is technically mandated; CodeQL requires initialization prior to the build so that compiler invocations can be monitored, and the analysis step must follow build completion~\cite{codeql_overview}. Consequently, the most reproducible workflow pattern in the dataset is not a build-and-test pipeline but a security analysis pipeline, consistent with GitHub's promotion of CodeQL scanning through default repository setup and security advisory notifications~\cite{codeql_overview}. Technically enforced structural templates thus produce more consistent adoption across projects and languages than voluntarily configured best practices.

\smallskip\noindent\textbf{Common pipeline opening patterns.}
Table~\ref{tab:prefixes} shows the most common 2-to-4-step prefixes.
The 2-step prefix \textit{Checkout $\rightarrow$ Setup Runtime
Environment} alone accounts for 39.5\% of all workflows, making it
the single most recognizable pattern in the dataset.
The 3-step extension that adds \textit{Install Project
Dependencies} reaches 20.9\%, and the 4-step variant that further
adds \textit{Execute Tests} or \textit{Run Static Analysis} each
clear the 5\% threshold, at 5.5\% and 6.3\% respectively.

\smallskip\noindent\textbf{Opening vs.\ Closing Variability.} GHA workflow openings converge around a recognizable pattern but conclusions remain highly variable.
The pipeline's conclusion is considerably more variable than its opening; only 5.5\% of workflows open with the complete four-step canonical sequence through to test execution. Across the sample, \textit{Upload Build Artifacts} (25.6\%) and \textit{Run Static Analysis} (22.0\%) are nearly as prevalent as \textit{Execute Tests} (27.8\%), indicating that in practice, workflows serve artifact-production and quality-checking purposes as frequently as they serve verification purposes. While CI pipeline entry points converge around a recognizable norm, pipeline conclusions remain highly project specific and do not consistently reflect the build-and-test model that GitHub's documentation presents as the canonical CI workflow~\cite{github_build_test_code}.
 
\begin{table}[ht]
  \centering
  \caption{Most frequent sequence prefixes (global, top-3 per
           length). Percentages relative to \(N=382\).}
  \label{tab:prefixes}
  \small
  \setlength{\tabcolsep}{4pt}
  \begin{tabular}{clrr}
    \toprule
    \textbf{Len} & \textbf{Prefix} & \textbf{Cnt} & \textbf{\%} \\
    \midrule
    2 & Checkout $\to$ Setup Runtime Env.           & 151 & 39.5 \\
    2 & Checkout $\to$ Init.\ Security Analysis     &  25 &  6.5 \\
    2 & Checkout $\to$ Setup Build Environment      &  22 &  5.8 \\
    \midrule
    3 & Checkout $\to$ Setup Runtime $\to$ Install Deps & 80 & 20.9 \\
    3 & Checkout $\to$ Init.\ Security $\to$ Build  &  22 &  5.8 \\
    3 & Checkout $\to$ Setup Runtime $\to$ Build    &  16 &  4.2 \\
    \midrule
    4 & Checkout $\to$ Setup $\to$ Install $\to$ Static Analysis & 24 & 6.3 \\
    4 & Checkout $\to$ Init.\ Security $\to$ Build $\to$ Sec.\ Analysis & 22 & 5.8 \\
    4 & Checkout $\to$ Setup $\to$ Install $\to$ Tests & 21 & 5.5 \\
    \bottomrule
  \end{tabular}
\end{table}
 
\smallskip\noindent\textbf{Common Pipeline Closing Patterns.}
The most common 2-step suffix, \textit{Build Project $\rightarrow$
Perform Security Analysis}, appears in 7.6\% of workflows, and the
3-step variant \textit{Initialize Security Analysis $\rightarrow$
Build Project $\rightarrow$ Perform Security Analysis} in 7.3\%,
indicating that security analysis is consistently observed at the pipeline's conclusion rather than prior to build execution.
 
\smallskip\noindent\textbf{Step transitions.}
The single most frequent consecutive pair globally is
\textit{Checkout Repository $\rightarrow$ Setup Runtime
Environment} (160 occurrences, 9.53\% of all transitions), followed
by \textit{Setup Runtime Environment $\rightarrow$ Install Project
Dependencies} (103 occurrences, 6.13\%). Together these two
transitions form the backbone of the canonical CI opening.
The transition \textit{Build Project $\rightarrow$ Upload Build
Artifacts} (1.85\%) and \textit{Execute Tests $\rightarrow$ Upload
Coverage} (0.77\%) mark the end of the productive pipeline body,
while \textit{Initialize Security Analysis $\rightarrow$ Build
Project} (1.67\%) and \textit{Build Project $\rightarrow$ Perform
Security Analysis} (1.73\%) together constitute the CodeQL
sub-pattern noted above.

\smallskip\noindent\textbf{Pipeline Closure.} Security analysis consistently closes the pipeline rather than opening it.
Both the suffix analysis and the step-transition data place security scanning at the end of the productive pipeline body, after the build step rather than before it. This confirms that the CodeQL pattern's fixed ordering, noted above, is a dataset-wide convention, not an artifact of the single dominant full sequence.

\smallskip\noindent\textbf{Language-Specific Sequence Patterns.} 
Table~\ref{tab:lang_struct} summarizes structural metrics per language, and Table~\ref{tab:lang_practices} summarizes adoption and practice metrics per language and reveals marked inter-language differences.
 
\begin{table*}[ht]
\centering
\caption{Structural and completeness metrics (\(N=382\)).}
\label{tab:lang_struct}
\small
\setlength{\tabcolsep}{4pt}
\begin{tabular}{lrrrrrrrr}
\toprule
\textbf{Lang} & \textbf{n} &
\textbf{Avg} & \textbf{Med.} &
\textbf{\%Ck} & \textbf{\%Setup} &
\textbf{\%Build} & \textbf{\%Test} &
\textbf{Compl.} \\
\midrule
C++    &  84 & 5.98 & 5.5 & 98.81 & 67.86 & 79.76 & 29.76 & 0.69  \\
Java   &  76 & 4.75 & 4.0 & 80.26 & 67.11 & 63.16 & 14.47 & 0.562 \\
Python & 222 & 5.40 & 5.0 & 92.79 & 70.72 & 47.30 & 31.53 & 0.606 \\
\midrule
Global & 382 & 5.40 & 5.0 & 91.62 & 69.37 & 57.59 & 27.75 & 0.616 \\
\bottomrule
\end{tabular}
\end{table*}

\begin{table*}[ht]
\centering
\caption{Adoption and practice metrics (\(N=382\)).}
\label{tab:lang_practices}
\small
\setlength{\tabcolsep}{4pt}
\begin{tabular}{lrrrrrr}
\textbf{\%Cache} & \textbf{\%Reuse} &
\textbf{\%SHA} & \textbf{\%Tag} &
\textbf{Cmn} \\
\midrule
C++    & 79.76 & 16.67 & 2.38 &  7.14 & 95.24 & 0 \\
Java   & 69.74 & 21.05 & 3.95 & 15.79 & 84.21 & 2 \\
Python & 86.04 &  7.21 & 3.15 &  2.70 & 96.40 & 2 \\
\midrule
Global & 81.41 & 12.04 & 3.14 &  6.28 & 93.72 & 1 \\
\bottomrule
\end{tabular}
\end{table*}

We observe that C++ workflows have the highest \textit{Build Project} rate (79.76\%)
and the longest average sequences (5.98 steps, median 5.5),
reflecting the compile-intensive nature of the language and the
tendency to run builds across multiple platform configurations.
Despite having the highest checkout \emph{presence} rate (98.81\%),
only 79.76\% of C++ workflows \emph{start} with checkout, because
workflows frequently open with environment-setup or cache-restore
steps placed before checkout to accelerate toolchain initialization
(e.g.\ pre-installed compilers, \texttt{cache} warm-up).
No C++ full sequence reaches the 5\% repetition threshold; the
most frequent is \textit{Checkout $\rightarrow$ Run Static Analysis}
(3.57\%), confirming that build-system fragmentation (CMake, Make,
Meson, MSBuild, Bazel) and platform-matrix strategies prevent
convergence on any single pipeline shape. Cache adoption is the
highest of the three languages (16.67\%), consistent with the
long compilation times that make caching high-value.

Java presents the most unusual structural profile: the lowest
checkout-presence rate (80.26\%), the lowest starts with checkout
rate (69.74\%),meaning nearly \emph{one in three} Java workflows
does not begin with source retrieval and by far the lowest test
rate (14.47\%), yielding the lowest completeness score (0.562).
These figures reflect Java's build tool integration: \texttt{mvn
verify} and \texttt{gradle build} embed compilation, testing, and
packaging into a single invocation, so a workflow containing only a
build construct may be executing tests invisibly to static analysis.
The maximum observed sequence length of 26 steps is the longest in
the dataset, belonging to a workflow that chains matrix jobs, artifact
exchange, and release publishing. Java has the highest cache adoption
(21.05\%), reflecting Maven's and Gradle's substantial benefit from
local repository caching~\cite{github_caching_dependencies}, and the
highest SHA-pinning rate (15.79\%), consistent with the enterprise
security practices prevalent in the Java community. Its two common
sequences are the CodeQL scanning pipeline (6.58\%) and
\textit{Checkout $\rightarrow$ Setup $\rightarrow$ Build} (5.3\%).
 
Python has the highest starts with checkout rate (86.04\%) and the
highest test rate (31.53\%), making it the closest language to the
canonical CI template. Its two common full sequences are the CodeQL
pattern (7.21\%) and \textit{Checkout $\rightarrow$ Setup
$\rightarrow$ Install $\rightarrow$ Test} (5.4\%), confirming that
Python projects have converged on two reproducible pipeline shapes
to a degree neither C++ nor Java has matched. Python also records the highest scheduled-workflow rate (25.23\%) and workflow-dispatch rate (27.93\%) among the three languages, suggesting a broader automation culture that extends GHA use beyond event driven CI. Scheduled workflows in Python-based repositories are consistent with activities such as periodic dependency audits and nightly documentation builds, while workflow dispatch triggers are commonly associated with manually controlled release pipelines~\cite{kinsman2021software}. Despite this operational sophistication, Python has the lowest cache adoption (7.21\%) and the lowest SHA-pinning rate (2.70\%), creating tension between automation breadth and security hygiene in the Python CI ecosystem.

\smallskip\noindent\textbf{Language-Specific Convergence.} Python workflows converge on reproducible pipeline shapes to a degree neither C++ nor Java matches.
Python is the only language with two full sequences clearing the 5\% commonality threshold, and it posts the highest checkout-start rate and test rate of the three languages. C++, by contrast, has no full sequence reaching that threshold at all, and Java posts the lowest completeness score (0.562) of the three, consistent with its build tools embedding test and package steps invisibly to static analysis (consistent with the pipeline-closing pattern above and the RQ1 findings on Java's action composition). Convergence on a shared pipeline shape is therefore a language-specific outcome, not a property of GHA itself.

\smallskip\noindent\textbf{Caching Adoption.} Dependency caching adoption is low overall but follows a language-stratified pattern consistent with toolchain overhead.
Java leads caching adoption (21.05\%), followed by C++ (16.67\%), with Python trailing (7.21\%) despite caching's documented impact on pipeline execution time~\cite{github_caching_dependencies}. This matches the heavier compilation and dependency-resolution overhead of Java and C++ toolchains~\cite{maven_lifecycle_intro,gradle_java_plugin}, where caching pays off more than in Python, whose shorter install times make the same configuration overhead less worthwhile. The step transition analysis shows that when caching is configured it sits immediately after checkout, among the top 20 overall step transitions, treated as an upfront configuration concern rather than a later optimization. This adoption rate measures explicit cache steps identified during open coding, whereas RQ3 rule EFF-01 applies only to workflows whose YAML contains recognizable install commands and flags the absence of a corresponding cache configuration; the two metrics differ in scope and are not direct complements.

\smallskip\noindent\textbf{Statistical Significance and Effect Sizes.}
Kruskal-Wallis testing detected statistically significant cross-language differences for five of eight continuous structural metrics ($p < 0.05$), including sequence length ($H = 10.4$, $p = 0.005$), total step count ($H = 13.1$, $p = 0.001$), and completeness score ($H = 7.6$, $p = 0.023$). Table~\ref{tab:rq2_stats} summarizes pairwise comparisons with \emph{small} or greater effect sizes. C++ workflows are significantly longer and step-heavier than Java workflows (sequence length: medians 5.5 vs.\ 4.0, $\delta = 0.278$, $p = 0.002$; total steps: medians 8 vs.\ 5, $\delta = 0.321$, $p < 0.001$). Proportion-based tests confirmed that build-step presence differs significantly across all three language pairs (Fisher's exact, $p < 0.05$ in each case; Cohen's $h = 0.32$ to $0.69$), as do explicit test-step rates between C++ and Java (29.8\% vs.\ 14.5\%, $h = 0.37$, $p = 0.024$) and between Java and Python (14.5\% vs.\ 31.5\%, $h = -0.41$, $p = 0.004$). Cache adoption is significantly lower in Python than in Java (7.2\% vs.\ 21.1\%, $h = 0.41$, $p = 0.002$) and C++ (7.2\% vs.\ 16.7\%, $h = 0.30$, $p = 0.018$), while SHA pinning is significantly more common in Java than Python (15.8\% vs.\ 2.7\%, $h = 0.49$, $p < 0.001$). As in RQ1, several omnibus tests reach significance with only negligible pairwise effects, underscoring the need to report effect sizes alongside $p$-values in stratified sample analyses.

\smallskip\noindent\textbf{Dominant Effects.} The largest, most consistent effects concentrate in build-versus-test presence and toolchain-driven practices, not raw sequence length.
The single largest effect in this RQ is the build-step presence gap between C++ and Python ($h = 0.69$, large), not a size metric, indicating that pipeline-stage presence, not length, most reliably separates languages in this sample.

\begin{table}[t]
  \centering
  \caption{Selected pairwise tests for RQ2 metrics with small or greater effect sizes ($N=382$).}
  \label{tab:rq2_stats}
  \small
  \setlength{\tabcolsep}{3pt}
  \begin{tabular}{llccccc}
    \toprule
    \textbf{Metric} & \textbf{Comparison}
      & \textbf{Val\textsubscript{1}}
      & \textbf{Val\textsubscript{2}}
      & \textbf{Effect}
      & \textbf{Size}
      & \textbf{$p$} \\
    \midrule
    Sequence length   & C++ vs Java   & 5.5 & 4.0 & $\delta{=}0.278$ & small & 0.002 \\
    Total steps       & C++ vs Java   & 8   & 5   & $\delta{=}0.321$ & small & $<0.001$ \\
    Total steps       & Java vs Python& 5   & 5   & $\delta{=}{-}0.160$ & small & 0.036 \\
    Completeness      & C++ vs Java   & 0.75& 0.75& $\delta{=}0.189$ & small & 0.029 \\
    Build presence    & C++ vs Python & 79.8\% & 47.3\% & $h{=}0.69$ & large & $<0.001$ \\
    Test presence     & Java vs Python& 14.5\% & 31.5\% & $h{=}{-}0.41$ & small & 0.004 \\
    Cache adoption    & Java vs Python& 21.1\% & 7.2\%  & $h{=}0.41$ & small & 0.002 \\
    SHA pinning       & Java vs Python& 15.8\% & 2.7\%  & $h{=}0.49$ & small & $<0.001$ \\
    \bottomrule
  \end{tabular}
\end{table}

\smallskip\noindent\textbf{Structural Workflow Deviations.} 
Table~\ref{tab:absence} reports eight canonical absence patterns
across the full dataset. \textit{Build without Test}, 42.1\% of workflows
include a build step but no automated test step, is the most common deviation from
the canonical CI pipeline recommended by GitHub
documentation~\cite{github_actions_docs}.
A further 22.8\% of workflows perform only a checkout with no
subsequent build or test; these represent administrative, labeling,
notification, or release management pipelines rather than
build and verify workflows.

\begin{table}[ht]
  \centering
  \caption{Category absence patterns (global, \(N=382\)).}
  \label{tab:absence}
  \small
  \begin{tabular}{lrr}
    \toprule
    \textbf{Pattern} & \textbf{Count} & \textbf{\%} \\
    \midrule
    Build without Test        & 161 & 42.1 \\
    Checkout only             &  87 & 22.8 \\
    Build without Setup       &  50 & 13.1 \\
    Test without Build        &  47 & 12.3 \\
    No Checkout, No Setup     &  32 &  8.4 \\
    Test without Setup        &  11 &  2.9 \\
    Test without Checkout     &   3 &  0.8 \\
    Build without Checkout    &   2 &  0.5 \\
    \bottomrule
  \end{tabular}
\end{table}

\smallskip\noindent\textbf{Build vs.\ Test Gap.} Explicit test steps are significantly less prevalent than build steps across all three languages, with the gap most pronounced in Java.
The gap between build and test adoption (44.2\% vs.\ 27.8\%, 16.4~pp) is most pronounced in Java, where 63.2\% of workflows include a build step yet only 14.5\% include an explicit test step, a 48.7~pp difference. This gap is attributable to Java build tools such as \texttt{mvn verify}~\cite{maven_lifecycle_intro} and \texttt{gradle build}~\cite{gradle_java_plugin}, which embed test execution within a single build invocation, making it invisible to static workflow analysis. In Python the gap narrows (47.3\% vs.\ 31.5\%, 15.8~pp), consistent with the language's interpreted nature requiring no compilation step. In C++ build steps dominate most absolutely (79.8\% vs.\ 29.8\%, 50~pp), reflecting the compilation-intensive nature of native code development rather than an absence of testing practice. Explicit test step presence is therefore an unreliable proxy for CI pipeline completeness, particularly for compiled language projects where testing may be embedded within build tool invocations or distributed across separate workflow files.

\smallskip\noindent\textbf{Workflows Not Starting with Checkout.} 
Of 382 workflows, \textbf{71 (18.6\%)} do not begin with
\textit{Checkout Repository} as their first step.
Table~\ref{tab:firstnc} lists the most common alternative first
steps.
 
\begin{table}[ht]
  \centering
  \caption{Most frequent first steps among non-checkout-first
           workflows (\(n=71\)).}
  \label{tab:firstnc}
  \small
  \begin{tabular}{lrr}
    \toprule
    \textbf{First Step} & \textbf{Count} & \textbf{\% of all} \\
    \midrule
    Manage Workflow Automation  & 10 & 2.6 \\
    Setup Runtime Environment   &  7 & 1.8 \\
    Publish Release             &  6 & 1.6 \\
    Execute Automation Script   &  4 & 1.0 \\
    Inspect Runtime Environment &  3 & 0.8 \\
    Install Project Dependencies&  3 & 0.8 \\
    Setup Build Environment     &  3 & 0.8 \\
    Extract Dependency Metadata &  3 & 0.8 \\
    Configure Build Profile     &  2 & 0.5 \\
    Send Notification           &  2 & 0.5 \\
    \bottomrule
  \end{tabular}
\end{table}

\smallskip\noindent\textbf{Entry Point Deviations.} Checkout is the near-universal entry point, yet deviations are most pronounced in Java.
The checkout step is the documented entry point for GHA workflows~\cite{github_actions_docs}, and 81.4\% of the 382 analyzed workflows conform to this norm. Java workflows show the sharpest deviation at 30.3\%, compared to 20.2\% in C++ and 14.0\% in Python. Among the Java workflows examined, this pattern is driven by a cluster of administrative workflows, including label managers, release publishers, and Dependabot automation pipelines~\cite{github_dependabot_quickstart}, that interact exclusively with GitHub's metadata APIs~\cite{github_event_types} and require no source code retrieval. These workflows are intentionally structured and reflect GHA's documented role as a general purpose automation platform beyond CI~\cite{kinsman2021software}. Any tool or study that treats checkout-first position as a fixed structural rule will misclassify nearly one in three Java workflows in this sample, underscoring the need for language- and purpose-aware assumptions in GHA research and tooling.
 
\smallskip\noindent\textbf{Trigger Usage and Construct Co-occurrence.} 
Push events are the most common trigger (74.4\%), followed by pull
request (62.0\%), workflow dispatch (24.9\%), and schedule
(19.9\%). Push-triggered workflows are more likely to include a
build step (+30.8 percentage points above non-push workflows) and
a setup step (+23.3 pp), suggesting they are the primary vehicle
for build verification. Scheduled workflows, by contrast, show
substantially lower rates of both setup ($-$38.9~pp) and checkout
($-$9.3~pp), consistent with their use for periodic maintenance
tasks such as dependency audits, artifact publication, and
repository housekeeping that do not require a full build
environment. Workflow dispatch triggers show the highest association
with reusable workflows (+9.8~pp), aligning with their role as
manually invoked or orchestrated meta-pipelines.

\smallskip\noindent\textbf{Trigger-Stage Correlation.} Trigger type is a strong predictor of which pipeline stages a workflow contains.
Push and pull-request triggers correlate with full build verification, scheduled triggers correlate with lightweight maintenance tasks that skip setup and checkout entirely, and workflow-dispatch triggers correlate with reusable-workflow usage. Trigger type is therefore not incidental metadata; it signals the functional category of the workflow and could serve as a low-cost heuristic for classifying workflow purpose without inspecting step content.
 
\smallskip\noindent\textbf{Modularity and Caching Adoption.} 
GHA provides two primary mechanisms for workflow
reuse~\cite{github_reusing_workflows}: reusable workflows and
composite actions. Adoption remains low: only 3.1\% of workflows
use reusable workflows and 0.3\% use composite actions.
Workflows that \emph{do} employ reusable workflows exhibit slightly
lower completeness scores (0.48 vs.\ 0.62 for non-users), likely
because they delegate build or test logic to the called workflow
rather than embedding it directly, making those constructs
invisible to static analysis of the caller.
Dependency caching (\textit{actions/cache}) is used in only 12.0\%
of workflows, surprisingly low given its direct impact on
pipeline execution time~\cite{github_caching_dependencies}.

\smallskip\noindent\textbf{Modularity Blind Spot.} Modularity adoption remains low and introduces a measurement blind spot that depresses completeness scores.
GHA's modularity mechanisms have not yet gained widespread uptake among the open-source projects in this sample, despite being available since 2021~\cite{github_reusing_workflows}. Because reusable workflows delegate steps to sub-workflows, those steps are statically invisible to the calling workflow's analysis, a measurement limitation, not a genuine quality deficit. It does show that completeness scores are unreliable proxies for pipeline quality in modular workflows, and that modularity adoption should be reported separately from construct-presence metrics.

\smallskip\noindent\textbf{Version Pinning Practices}
 Table~\ref{tab:pinning} shows version pinning practices by
language. Tag-based pinning (e.g.\ \texttt{@v3}) dominates
across all three languages (84\% to 96\%), while commit-SHA pinning,
the practice recommended by GitHub security guidelines~\cite{github_security_hardening} to prevent supply-chain
attacks via mutable tags, is used in only 6.3\% of workflows
globally. Java projects show the highest SHA adoption (15.79\%),
while Python projects show the lowest (2.70\%). Branch based
references, which are the least stable pinning strategy, are most
prevalent in Python (15.8\%), potentially reflecting looser
dependency hygiene in Python's CI ecosystem or greater reliance on
development branch actions.
 
\begin{table}[ht]
  \centering
  \caption{Version pinning practices by language (\% of workflows
           with at least one action using that pinning strategy).}
  \label{tab:pinning}
  \small
  \begin{tabular}{lrrrrrr}
    \toprule
    \textbf{Lang} & \textbf{n} & \textbf{SHA\%} &
    \textbf{Tag\%} & \textbf{Branch\%} & \textbf{Unpinned\%} \\
    \midrule
    Java   & 76  & 15.79 & 84.21 &  5.3 & 6.58 \\
    C++    & 84  &  7.14 & 95.24 & 10.7 & 4.76 \\
    Python & 222 &  2.70 & 96.40 & 15.8 & 2.25 \\
    \midrule
    Global & 382 &  6.3 & 93.7 & 12.6 & 3.7 \\
    \bottomrule
  \end{tabular}
\end{table}

\smallskip\noindent\textbf{Pinning Practices.} Tag-based pinning dominates across all languages while SHA pinning, the only tamper-resistant strategy, remains critically underadopted.
The dominant pinning strategy is technically insufficient to prevent supply-chain compromise: only SHA pinning is immune to tag-mutation attacks~\cite{github_security_hardening}. Since Python accounts for 58.1\% of the workflows in this sample, the majority of analyzed workflows operate with a pinning strategy vulnerable to tag-mutation supply-chain attacks, as demonstrated by the 2025 \textit{tj-actions/changed-files} incident in which tagged action releases were silently rewritten to exfiltrate secrets~\cite{stepsecurity_github}.
  
\vspace{-6pt}
\begin{rqbox}
\textbf{RQ\textsubscript{2} Summary.}
CI pipelines converge on a canonical two-step opening (checkout, then environment setup) but diverge sharply after that: the only globally reproducible full sequence is a CodeQL security-scanning pipeline, not a build-and-test one, and 42.1\% of workflows build without testing, most acutely in Java. Modularity mechanisms and SHA pinning remain rare, and dependency caching adoption tracks each language's build toolchain overhead.
\end{rqbox}

\subsection{\textbf{RQ3: To what extent do GHA workflows comply with official documentation and best practices?}}

\subsubsection{Motivation}

GHA documentation and best-practice guidelines aim to make CI easier to adopt while reducing misconfigurations that affect reliability, reproducibility, and security. In principle, if developers follow this guidance, it should act as a normalising force across open-source workflows. However, prior studies on CI configuration smells and security practices suggest that workflows often deviate from these recommendations~\cite{koishybayev2022characterizing, gallaba2020tse, hilton2016usage}, raising questions about how well the documentation aligns with real-world use.
These deviations are not random. They may reflect usability gaps, where recommended patterns are hard to discover or apply in practice; ecosystem specific constraints, where guidance does not translate cleanly across languages and toolchains (e.g., C++ or Python environments); or mismatches in priorities, where developers defer certain practices until failures expose their importance.
Motivated by this gap between prescribed guidance and observed practice, this study examines which best practices are violated, how frequently these violations occur, and how they vary between ecosystems. By assessing adherence at scale, the study aims to identify where documentation falls short in practice and where it may need to evolve to better support developers in real CI environments~\cite{hassan2008road}

\subsubsection{Approach}

\smallskip\noindent\textbf{Compliance Rules.}
We analyzed \textbf{382 GitHub Actions workflow files} spanning three programming language ecosystems: \textbf{Python} ($n = 222$), \textbf{C++} ($n = 84$), and \textbf{Java} ($n = 76$), and operationalized compliance using ten rules derived from (i)~official GitHub Actions documentation~\cite{github_actions_docs, github_security_hardening, github_token} and (ii)~established CI best-practice sources~\cite{koishybayev2022characterizing, decan2022use}. Each rule captures a structural or semantic property that is statically detectable in YAML and directly grounded in documented guidance. The rules are grouped into five violation categories: \textbf{Security} (SEC-01 to SEC-03), \textbf{Reliability} (REL-01 and REL-02), \textbf{Versioning} (VER-01), \textbf{Efficiency} (EFF-01 and EFF-02), \textbf{Maintainability} (MAINT-01), with an additional rule (WA-01) capturing privileged workflow triggers as a \emph{Workflow Activator} concern.
We treated any deviation from the GHA documented guidance as a bad practice, consistent with prior CI configuration studies~\cite{gallaba2020tse, vassallo2016continuous, zampetti2017open, khatami2024catching, zampetti2020empirical, bouzenia2024resource}. Each rule is assigned an integer severity weight on a scale of 1 to 5 based on the potential consequence of non-compliance. Rules whose violation can directly enable a security compromise, secret interpolation (SEC-01), missing permission scoping (SEC-02), unpinned third-party actions (SEC-03), and privileged fork triggers (WA-01), are assigned the maximum weight of \textbf{5}, consistent with their classification as critical risks in the GHA security hardening documentation~\cite{github_security_hardening} and in Koishybayev et al.~\cite{koishybayev2022characterizing}. Silent failure masking (REL-02) is weighted \textbf{4}, as it actively conceals build failures without causing direct security harm. The remaining five rules, missing timeouts (REL-01), mutable runner labels (VER-01), absent caching (EFF-01, EFF-02), and duplicated logic (MAINT-01), are each weighted \textbf{3}, as their violations degrade reliability, reproducibility, or efficiency without introducing an immediate security consequence.

\smallskip\noindent\textbf{Compliance Scoring.}
Using this rule set, we perform a comparative analysis across projects and languages to quantify adherence, examine its distribution across ecosystems, and relate violations to project characteristics. The rule detection logic is informed by the open coding and automated YAML parsing conducted in RQ2. Structural patterns identified in RQ2, such as job dependencies, trigger configurations, and step composition, provide contextual signals for specific rules. For example, MAINT-01 targets duplicated pipeline logic, supported by the low reusable workflow adoption rate (3.1\%) observed in RQ2, while WA-01 flags workflows using \texttt{pull\_request\_target} triggers that grant elevated permissions to forked pull requests~\cite{github_security_hardening}.
We compute compliance as a weighted average, expressing the proportion of satisfied rules weighted by their severity and assigning each workflow a score between 0 and 1:

\begin{equation*}
  \text{Compliance Score} = \frac{\text{Rules passed (weighted by severity)}}
                                 {\text{All rules (weighted by severity)}}
\end{equation*}

A score of \textbf{1.0} means the workflow passes all ten rules; a score of \textbf{0.0} means it fails all of them at the highest severity.
Recognizing that not all deviations are equally problematic in every project context~\cite{zampetti2017open,palomba2018diffuseness}, each violation instance was annotated with a \emph{justifiability assessment}: \textit{NOT justifiable}, \textit{POSSIBLY justifiable}, or \textit{context-dependent}. For example, the absence of dependency caching (EFF-01) may be intentional in a workflow explicitly designed to audit clean installs, whereas the omission of a \texttt{permissions:} block (SEC-02) is never justifiable unless an organization-level read-only token policy is externally enforced. These annotations enrich the binary violation signal and allow downstream analysis to distinguish configuration debt from deliberate trade-offs~\cite{zampetti2017open}.

\smallskip\noindent\textbf{Statistical Analysis.}
The compliance analysis is conducted comparatively across the three language ecosystems in our dataset (Python, C++, Java), enabling us to test whether violation patterns are language-agnostic or ecosystem-specific. Cross-language differences in compliance score, violation count, and per-rule violation rates were tested using the same framework as RQ2: Kruskal-Wallis and Mann-Whitney $U$ tests with Cliff's $\delta$ for continuous outcomes, and chi-square and Fisher's exact tests with Cohen's $h$ for binary rule violations. We further examine whether project-level characteristics surfaced in RQ1 (e.g., workflow complexity, step count, trigger diversity) are associated with lower compliance scores. Bad practices are ranked using the composite risk score defined as $\text{risk} = (\text{prevalence} \times \text{severity}) / 5$, providing a prioritized ordering that accounts for both how often a violation occurs and how harmful it is when it does, in alignment with prior CI smell analysis work~\cite{zampetti2017open, vassallo2016continuous}.

\subsubsection{Findings}

\smallskip\noindent\textbf{Global Compliance Scores.} 
The dataset-wide compliance scores are summarized in
Table~\ref{tab:rq3_global_stats}. The mean compliance score across all 382 workflows is $\bar{x} = 0.72$ ($\sigma = 0.10$), with a median of 0.72, indicating a moderately right-skewed distribution truncated at perfect compliance. Only \textbf{1 workflow} (0.3\%) achieved perfect compliance, the \texttt{ci.yml} workflow from the C++ project; \texttt{uxlfoundation/oneDAL}, with a further 7 workflows (1.8\%) scoring 0.92, the next highest attainable score given the ten-rule structure. At the other end, no workflow scored below 0.33, and only 7 workflows (1.8\%) fell below 0.50, indicating that even the least compliant workflows still satisfied a meaningful portion
of the rules. This points to a baseline level of compliance awareness across all three ecosystems, with most workflows falling short not through ignorance of all practices, but through consistent omission of a small cluster of specific rules. On average, each workflow violated 3 out of 10 rules, with a weighted severity average of 10.92 out of a possible 39. The sole perfectly compliant workflow belonged to a mature, foundation-backed C++ project (\texttt{uxlfoundation/oneDAL}); whether 
organizational maturity is a contributing factor warrants further 
investigation beyond the scope of this study.
 
\begin{table}[ht]
  \centering
  \caption{Overall and per-language compliance statistics ($N = 382$ workflows).}
  \label{tab:rq3_global_stats}
  \renewcommand{\arraystretch}{1.25}
  \begin{tabular}{lcccccc}
    \toprule
    \textbf{Ecosystem} & \textbf{$n$} & \textbf{Mean score} & \textbf{Median} & \textbf{$\sigma$} & \textbf{Min} & \textbf{Max} \\
    \midrule
    Python  & 222 & 0.706 & 0.718 & 0.105 & 0.333 & 0.923 \\
    C++     & 84  & 0.720 & 0.718 & 0.095 & 0.513 & 1.000 \\
    Java    & 76  & 0.762 & 0.795 & 0.101 & 0.410 & 0.923 \\
    \midrule
    \textbf{All} & \textbf{382} & \textbf{0.720} & \textbf{0.718} & \textbf{0.101} & \textbf{0.333} & \textbf{1.000} \\
    \bottomrule
  \end{tabular}
\end{table}
 
Java workflows exhibit the highest mean compliance score (0.762) and the lowest mean number of violations per file (2.61), driven largely by near-zero EFF-01 violation rates among Java workflows (1.3\%). Python workflows recorded the lowest mean score (0.706) and the highest violation density (3.27 per workflow), with EFF-01 as the primary differentiator: 44.6\% of Python workflows violate EFF-01, compared to 20.2\% for C++ and 1.3\% for Java. Because EFF-01 is triggered only when install commands are detected in the YAML (128 of 222 Python workflows, 31 of 84 C++, and 4 of 76 Java), the rates above are computed against the full per-language sample; restricted to install-bearing workflows only, where the rule substantively applies, the Python figure rises to 77.3\%.
Python also records the highest mutable runner label rate (VER-01: 85.6\%), suggesting that reproducibility concerns receive less attention in that ecosystem. C++ sits between the two, but leads in missing permissions blocks (SEC-02: 83.3\%), likely reflecting the lower prevalence of secrets-dependent workflows in systems level projects where the risk is less immediately visible to contributors.

\smallskip\noindent\textbf{Statistical Significance and Effect Sizes.}
Kruskal-Wallis tests confirm that compliance scores differ significantly across languages ($H = 19.1$, $p < 0.001$), as do total violation counts ($H = 25.8$, $p < 0.001$). Pairwise comparisons (Table~\ref{tab:rq3_stats}) show that Java workflows score significantly higher than Python (median 0.769 vs.\ 0.718, $\delta = 0.321$, $p < 0.001$) and C++ ($\delta = -0.283$, $p = 0.001$), while C++ and Python do not differ significantly ($p = 0.36$, negligible $\delta$). Rule-level Fisher's exact tests identify EFF-01 as the most language-stratified violation: Python workflows violate this rule far more often than Java (44.6\% vs.\ 1.3\%, $h = -1.23$, $p < 10^{-14}$) and C++ (44.6\% vs.\ 20.2\%, $h = -0.53$, $p < 0.001$). VER-01 violations are also significantly more prevalent in Python than C++ (85.6\% vs.\ 58.3\%, $h = -0.62$, $p < 0.001$), and SEC-02 violations are more prevalent in C++ than Java (83.3\% vs.\ 63.2\%, $h = 0.46$, $p = 0.004$).

\begin{table}[ht]
  \centering
  \caption{Selected pairwise compliance comparisons with small or greater effect sizes ($N=382$).}
  \label{tab:rq3_stats}
  \small
  \setlength{\tabcolsep}{4pt}
  \begin{tabular}{llccccc}
    \toprule
    \textbf{Outcome} & \textbf{Comparison}
      & \textbf{Med\textsubscript{1}}
      & \textbf{Med\textsubscript{2}}
      & \textbf{Effect}
      & \textbf{Size}
      & \textbf{$p$} \\
    \midrule
    Compliance score  & Java vs Python & 0.769 & 0.718 & $\delta{=}0.321$ & small & $<0.001$ \\
    Compliance score  & C++ vs Java    & 0.718 & 0.769 & $\delta{=}{-}0.283$ & small & 0.001 \\
    Violation count   & Java vs Python & 3     & 3     & $\delta{=}{-}0.367$ & medium & $<0.001$ \\
    EFF-01 violation  & Java vs Python & 1.3\% & 44.6\% & $h{=}{-}1.23$ & large & $<0.001$ \\
    VER-01 violation  & C++ vs Python  & 58.3\% & 85.6\% & $h{=}{-}0.62$ & medium & $<0.001$ \\
  \bottomrule
  \end{tabular}
\end{table}

\smallskip\noindent\textbf{Language Compliance Differences.} Compliance is moderate overall, with statistically detectable but modest language differences.
Java's higher mean score is consistent with the language's longer enterprise adoption history and the wider use of standardized build tooling (Maven/Gradle) that often ships with CI templates~\cite{hilton2016usage}. C++ and Python being statistically indistinguishable in overall compliance indicates that ecosystem-specific rule violations (EFF-01 in Python, SEC-02 in C++) offset one another at the aggregate score level.

\smallskip\noindent\textbf{Compliance Score Distribution.} Table~\ref{tab:rq3_distribution} shows the distribution of compliance scores across discrete bands. Nearly half of all workflows (45.0\%) fall in the $[0.70, 0.80)$ band, forming a pronounced mode across all three ecosystems. The lower tail ($< 0.60$) accounts for 12.0\% of the dataset (46 workflows), with broadly similar rates across languages: 13.1\% for Python, 11.9\% for C++, and 9.2\% for Java. At the upper end, only 8 workflows (2.1\%) scored $\geq 0.90$, five of which were Java projects, and just one, the \texttt{uxlfoundation/oneDAL} C++ workflow, achieved a perfect score of 1.0. This underlines that near-complete compliance is exceedingly rare regardless of language ecosystem.
 
\begin{table}[ht]
  \centering
  \caption{Distribution of compliance scores across the dataset.}
  \label{tab:rq3_distribution}
  \renewcommand{\arraystretch}{1.2}
  \begin{tabular}{lrr}
    \toprule
    \textbf{Score band} & \textbf{Count} & \textbf{\% of dataset} \\
    \midrule
    $[0.90,\ 1.00]$ & 8   & 2.1\%  \\
    $[0.80,\ 0.90)$ & 75  & 19.6\% \\
    $[0.70,\ 0.80)$ & 172 & 45.0\% \\
    $[0.60,\ 0.70)$ & 81  & 21.2\% \\
    $[0.50,\ 0.60)$ & 39  & 10.2\% \\
    $[0.00,\ 0.50)$ & 7   &  1.8\% \\
    \bottomrule
  \end{tabular}
\end{table}

\smallskip\noindent\textbf{Near-Perfect Compliance.} Near-perfect compliance is achievable through a single remaining violation.
Of the eight workflows scoring $\geq 0.90$, all seven that fell short of a perfect score missed exactly one rule. In every case, this violation was either \textbf{REL-01} (missing job timeout, 4 workflows) or \textbf{VER-01} (mutable runner label, 3 workflows); both are severity-3 rules. The gap from near-compliant to fully compliant is therefore narrow and actionable: correcting a single low-cost omission would have elevated these workflows to perfect compliance. This also confirms that the high-severity security rules (\textbf{SEC-01}, \textbf{SEC-03}) are not the limiting factors separating good from excellent workflows.
 
\smallskip\noindent\textbf{Violation Rates by Category.} Table~\ref{tab:rq3_categories} reports the proportion of workflows that trigger at least one violation in each category. \emph{Reliability} (96.3\%) and \emph{Versioning} (78.3\%) dominate, followed closely by \emph{Security} (72.5\%). Collectively, these three categories account for the vast majority of observed non-compliance, while \emph{Maintainability} is the least frequently violated (13.4\%).
 
\begin{table}[ht]
  \centering
  \caption{Proportion of workflows with at least one violation per category.}
  \label{tab:rq3_categories}
  \renewcommand{\arraystretch}{1.2}
  \begin{tabular}{lrrl}
    \toprule
    \textbf{Category} & \textbf{Workflows violated} & \textbf{\%} & \textbf{Primary rule} \\
    \midrule
    Reliability      & 368 & 96.3\% & REL-01 (Missing job timeout) \\
    Versioning       & 299 & 78.3\% & VER-01 (Mutable runner label) \\
    Security         & 277 & 72.5\% & SEC-02 (Missing permissions block) \\
    Efficiency       & 139 & 36.4\% & EFF-01 (No dependency caching) \\
    Maintainability  & 51  & 13.4\% & MAINT-01 (Duplicated workflow logic) \\
    \bottomrule
  \end{tabular}
  \vspace{-6pt}
\end{table}

\smallskip\noindent\textbf{Violation Category Concentration.} Non-compliance concentrates in reliability, versioning, and security, while maintainability is comparatively well handled.
Each of the three dominant categories corresponds to a single, low-effort field (a timeout bound, a pinned runner image, a permissions block), consistent with the broader pattern: non-compliance in this dataset is a problem of omission, not of varied mistakes.

\smallskip\noindent\textbf{Ranked Bad-Practice Patterns.} Table~\ref{tab:rq3_ranked} presents all ten compliance rules ranked by a composite \emph{risk score} defined as $\text{risk} = \text{prevalence} \times \text{severity} / 5$. This metric prioritizes patterns that combine high occurrence with high potential impact.
 
\begin{table}[ht]
  \centering
  \caption{Severity-weighted ranking of bad-practice patterns in GitHub Actions workflows ($N = 382$). 
  Prevalence denotes the percentage of workflows violating a rule. 
  Risk score is computed as $(\text{prevalence} \times \text{severity}) / 5$.}
  \label{tab:rq3_ranked}
  \renewcommand{\arraystretch}{1.2}
  \begin{tabularx}{\linewidth}{c l Y l c c c}
    \toprule
    \textbf{Rank} & \textbf{Rule} & \textbf{Description} & \textbf{Category} & \textbf{Sev.} & \textbf{Prev. (\%)} & \textbf{Risk} \\
    \midrule
    1  & SEC-02   & Missing permissions block                     & Security        & 5 & 71.2 & 3.56 \\
    2  & REL-01   & Missing job timeout                           & Reliability     & 3 & 96.3 & 2.89 \\
    3  & VER-01   & Mutable runner label                          & Versioning      & 3 & 78.3 & 2.35 \\
    4  & EFF-01   & No dependency caching                         & Efficiency      & 3 & 30.6 & 0.92 \\
    5  & MAINT-01 & Duplicated workflow logic                     & Maintainability & 3 & 13.4 & 0.40 \\
    6  & SEC-03   & Action pinned to branch/tag                & Security        & 5 &  5.8 & 0.29 \\
    7  & SEC-01   & Secret interpolation in \texttt{run} steps    & Security        & 5 &  4.7 & 0.24 \\
    8  & EFF-02   & Static/missing cache key                   & Efficiency      & 3 &  5.8 & 0.17 \\
    9  & REL-02   & Silent failure via \texttt{cont.-on-error} & Reliability     & 4 &  2.6 & 0.10 \\
    10 & WA-01    & Privileged trigger exposing forked code       & Workflow Activator & 5 & 0.0 & 0.00 \\
    \bottomrule
  \end{tabularx}
\end{table}
 
\textbf{The three highest risk patterns are elaborated below;}

\smallskip\noindent\textbf{SEC-02: Missing \texttt{permissions} block (risk: 3.56).} \textbf{71.2\%} of analyzed workflows omit an explicit \texttt{permissions:} declaration at either the workflow or job level. GitHub Actions assigns the \texttt{GITHUB\_TOKEN} \emph{write-all} access by default~\cite{github_token}, exposing repositories to unintended write operations or privilege escalation via a compromised third-party action. This finding corroborates Koishybayev et al.~\cite{koishybayev2022characterizing}, who reported 99.8\% over-privilege in a smaller earlier dataset, and indicates that despite ongoing documentation efforts, developers consistently overlook token scoping.
 
\smallskip\noindent\textbf{REL-01: Missing job timeout (risk: 2.89).}
An overwhelming \textbf{96.3\%} of workflows define no \texttt{timeout-minutes} bound on their jobs. GitHub's default of 360 minutes means that a hung dependency download, compilation step, or test suite can silently consume six hours of runner quota before failing. The pattern is language-agnostic: C++ link steps and Maven dependency resolution are equally susceptible to unbounded hangs. Although some long-running jobs may tolerate wider windows, the near-universal omission indicates a documentation gap rather than a deliberate configuration choice.
 
\smallskip\noindent\textbf{VER-01: Mutable runner label (risk: 2.35).}
We observe \textbf{78.3\%} of workflows use \texttt{ubuntu-latest} (or equivalent
\texttt{*-latest} labels) rather than a pinned image (e.g.
\texttt{ubuntu-22.04}). GitHub silently redirects \texttt{-latest} labels to newer images on their own schedule, which can transparently change the compiler toolchain, system library versions, or available interpreter runtimes between runs. This introduces non-reproducibility that mirrors the \emph{mutable tags} problem in Docker-based CI.

\smallskip\noindent\textbf{Omission vs.\ Commission.} Non-compliance is driven by omission, not commission.
The three highest-risk violations, missing \texttt{permissions} blocks, absent timeouts, and mutable runner labels, are all \emph{omissions} of recommended fields rather than the active misuse of dangerous constructs. This pattern implies that GHA's permissive defaults (write-all token, no timeout limit, latest runner aliasing)~\cite{github_token} function as traps: the safe behavior requires \emph{explicit opt-in}, yet the documentation does not surface these requirements prominently at the point of workflow authorship~\cite{koishybayev2022characterizing}. An overlapping subset of \textbf{68 workflows (17.8\%)} that violate SEC-02, REL-01, and VER-01 together also violate \textbf{EFF-01} (no dependency caching among install-bearing workflows). This points to a shared underlying cause, likely permissive platform defaults, rather than independent developer oversights: a developer who omits a \texttt{permissions:} block is often also unaware of the need to set \texttt{timeout-minutes} or pin a runner version.

\smallskip\noindent\textbf{Lower-risk patterns.}
EFF-01 (no dependency caching, 30.6\%) causes measurable runner-time waste but carries a lower severity (3) owing to partial justifiability; some workflows intentionally test fresh install paths. MAINT-01 (duplicated logic, 13.4\%) reflects underuse of GHA's reusable workflow and composite action primitives. Critically, the highest-severity security patterns SEC-01 (secret interpolation,
4.7\%) and SEC-03 (action pinned to branch, 5.8\%) occur rarely, and WA-01 (privileged fork trigger) registers \emph{zero} violations, indicating that developers do exercise caution around the most acute supply-chain attack vectors.

\smallskip\noindent\textbf{High-Severity Rarity.} High-severity security violations are rare but warrant monitoring.
Developers broadly avoid the most actively documented attack vectors~\cite{koishybayev2022characterizing}, yet \emph{structural} permission scoping (SEC-02) remains widely overlooked at 71.2\%. The most frequent rule combination overall is \textbf{SEC-02, REL-01, and VER-01}, occurring together in \textbf{196 workflows (51.3\%)}, reinforcing that non-compliance in this dataset is concentrated in a small cluster of omission-type rules rather than spread evenly across all ten.

\vspace{-6pt}
\begin{rqbox}
\textbf{RQ\textsubscript{3} Summary.}
Mean compliance sits at 0.72 across all three languages, with Java scoring modestly higher; differences are statistically detectable but not large. Non-compliance is dominated by omission, missing permissions, timeouts, and pinned runner versions, rather than active misuse, and over half of all workflows violate the same three rules together. High-severity security violations are rare, and the gap between near-perfect and perfect compliance typically comes down to a single missing field.
\end{rqbox}

\section{Discussion}
\label{sec:discussion}

\subsection{Implications for CI services}

    \vspace{3pt}
    \noindent \textit{\textbf{Limited Parallelism and High External Dependency Usage in CI Workflows.}} Eighty-five percent of workflows have a vertical dependency depth of 1 and 70\% contain no more than 70 lines of YAML, indicating that most open-source CI pipelines are structurally simple, consisting of a single job with sequentially executed steps. This is reinforced by a median job count of 1 across Python, C++, and Java, meaning CI platforms routinely provision runner infrastructure for pipelines that do not use available parallelism. C++ and Python workflows adopt matrix strategies at rates of 37.1\% and 34.3\% respectively, meaning a 3$\times$3 matrix multiplies runner consumption ninefold per trigger event across a substantial share of the dataset. Over 92.9\% of all workflows also invoke at least one external action from the GitHub Marketplace. Together these patterns present CI providers with a dual challenge: idle parallelism capacity in the majority of runs, and pervasive supply-chain exposure in nearly all of them.
    
    \vspace{3pt}
    \noindent \textit{\textbf{Adoption Gaps Reveal the Limits of Documentation-Driven Feature Promotion.}} The contrast between low voluntary adoption of reusable workflows (3.1\%) and dependency caching (12.0\%), and the comparatively consistent adoption of the CodeQL scanning pattern (5.8\%, the only globally common full sequence in our dataset), suggests that features requiring voluntary developer configuration are less likely to reach widespread adoption than those embedded in technically enforced default templates. Providers seeking to standardize practices such as caching, SHA pinning, and workflow reuse should consider proactive surfacing mechanisms such as inline configuration suggestions, repository health dashboards, and language aware default templates rather than relying on documentation alone. The 18.6\% of workflows that bypass repository checkout, and the substantial proportions using scheduled (19.9\%) and workflow-dispatch (24.9\%) triggers, confirm that a significant share of GHA usage falls outside the build-and-test paradigm; platform analytics and feature design that assume all workflows are CI pipelines will misrepresent actual platform usage. Finally, the language stratified patterns observed across C++, Java, and Python in build rates, caching adoption, and security practices indicate that pipeline health indicators and best-practice recommendations should be calibrated to language specific norms rather than applied uniformly.
    
    \vspace{3pt}
    \noindent \textit{\textbf{Default-driven, systematic non-compliance (in GitHub Actions).}} Non-compliance in GitHub Actions is largely systematic and driven by platform design rather than isolated developer error. The tightly clustered compliance scores reflect a consistent baseline of partially specified configurations, while the dominance of omission-based violations, particularly the frequent co-occurrence of SEC-02, REL-01, and VER-01, points to permissive defaults that neither require nor sufficiently surface essential fields. At the same time, the low incidence of high-severity issues suggests that developers generally avoid critical misuses, and the finding that near-perfect compliance is often blocked by a single missing configuration shows that the remaining gaps are both minor and easily addressable.
    Taken together, these results suggest that GitHub Actions and other CI platforms that similarly depend on developer opt-in for security-critical settings can improve compliance through targeted platform level interventions. Strengthening default configurations, prompting the completion of commonly omitted fields, and integrating lightweight best-practice checks directly into the authoring workflow would likely yield substantial improvements in configuration quality without increasing developer burden.

\subsection{Implications for Researchers}

    \vspace{3pt}
    \noindent \textit{\textbf{Statistical Significance Masks Minimal Practical Differences Across Workflow Metrics.}}
    Kruskal-Wallis tests confirmed statistically significant cross-language differences across all 18 metrics ($p < 0.05$ in every case, $p < 10^{-8}$ for 16 of 18), yet the majority of pairwise comparisons yielded only negligible Cliff's $\delta$ values below 0.147. Only seven comparisons crossed into small effect territory, and all seven involved C++ versus Java or Python on size and step heaviness metrics, or the external-action diversity ratio. Statistical significance, readily achieved at dataset scale, should therefore not be conflated with practical relevance in workflow complexity studies. The external-action diversity ratio produced the single largest effect size across all 18 metrics ($H = 765.7$, $p < 0.0001$, $\delta = -0.234$ for C++ vs Java), making it the most discriminating measure in this study. That this gap is rooted in Marketplace ecosystem maturity rather than developer intent underscores a broader methodological point: effect-size reporting is essential for distinguishing differences that are statistically detectable from those that are practically and structurally meaningful.
    
    \vspace{3pt}
    \noindent \textit{\textbf{Language Stratified Analysis Is Necessary to Avoid Conflating Structurally Distinct CI Conventions.}} Two assumptions common in GHA research, treating checkout-first position as a structural invariant and using explicit test-step presence as a proxy for CI completeness, fail to account for the diverse structural and toolchain conventions of different programming languages. The 30.3\% non-checkout-first rate in Java workflows and the tendency of compiled languages to embed testing within build tool invocations indicate that static workflow analysis may systematically underestimate the CI completeness of compiled-language projects. The contrast between step category frequency, which surfaces 21 common patterns at the 5\% threshold, and full sequence matching, which yields only a single globally common sequence, shows that category-level analysis is a substantially more productive unit for cross-project CI research than exact sequence matching. The measurement blind spot introduced by workflow modularity, where reusable workflows conceal steps from static analysis and produce artificially lower completeness scores (0.48 vs.\ 0.62 for non-modular workflows), indicates that construct presence counts alone are unreliable quality proxies for modular workflows. To avoid conflating structurally distinct CI conventions into misleading cross-language averages, future studies should treat language as a critical confounding variable and adopt stratified analytical frameworks that account for build tool semantics and modular workflow architecture.
    
    \vspace{3pt}
    \noindent \textit{\textbf{Systematic, omission-driven compliance as a modeling and generalization problem.}} CI non-compliance is not random but highly structured, characterized by tightly clustered compliance scores (mean 0.72) and dominated by omission-based violations, with SEC-02, REL-01, and VER-01 frequently co-occurring across more than half of the workflows. This consistency suggests that compliance should be studied not as isolated errors, but as a predictable outcome of underlying factors that remain unmodeled in this work. For researchers, this motivates a shift toward explanatory and predictive analyses that quantify the drivers of these patterns (e.g., project characteristics, template usage, or platform defaults), as well as investigations into their generalizability across CI ecosystems. Beyond this, the contrast between low rates of high-severity violations and widespread structural omissions highlights the need to move beyond syntactic rule checking toward approaches that assess whether workflows achieve their intended security and reproducibility guarantees in practice.
    
\subsection{Implications for Developers}

    \vspace{3pt}
    \noindent \textit{\textbf{Systematic Underuse of Reusable CI Constructs and Its Impact on Workflow Duplication.}} Reusable workflows are adopted by no more than 13.0\% of any language community, local custom actions by no more than 4.3\%, and deep job-dependency chains by no more than 5.3\% of workflows in the very high band, despite all three features being natively supported by the GitHub Actions platform. At the same time, the same build-test-publish logic is being independently reimplemented across thousands of repositories: C++ workflows, for instance, have a median of 6 steps and a 90th-percentile of 196 lines, with most of those steps consisting of raw \texttt{run:} shell blocks rather than reusable action invocations, as evidenced by C++'s external-action diversity ratio of only 0.35 compared to 0.53 for Java. Concretely, developers in C++ and Python communities are repeatedly writing and maintaining bespoke shell-scripted build pipelines instead of extracting shared, versioned workflow components, incurring duplicated maintenance burden at scale and forgoing the modularity and auditability that reusable workflows and local actions are designed to provide.
    
    \vspace{3pt}
    \noindent \textit{\textbf{The prevailing GHA usage patterns reveal a significant maturity gap in security, performance, and maintainability.}} While developers have successfully adopted GHA for basic automation, this maturity gap is evidenced by the high volume of workflows that build software without executing tests (42.1\%) and the near-universal reliance on mutable tag-based pinning (93.7\%), which offers a false sense of version security compared to the tamper-resistant SHA-pinning recommended by GitHub. The low adoption of dependency caching (12.0\%) and modular reuse mechanisms like reusable workflows (3.1\%) further suggests that most projects incur unnecessary costs in both execution time and technical debt. To capture the full value of the platform, developers must shift from "functional" configurations to "optimized" ones, prioritizing test integration for validation, SHA-pinning for supply-chain security, and caching and modularity for operational efficiency. This requires moving beyond initial setup to make these best practices default-on.
    
    \vspace{3pt}
    \noindent \textit{\textbf{Prioritize structural completeness for predictable workflow quality.}} The majority of non-compliance in GitHub Actions arises from the systematic omission of a few key fields, specifically SEC-02 (permissions scoping), REL-01 (job timeouts), and VER-01 (mutable runner labels), rather than active misuse. For developers, this implies that addressing these high-risk omissions first can yield immediate and measurable improvements in workflow reliability, security, and reproducibility. Practically, this means explicitly defining scoped \texttt{permissions:} blocks, setting \texttt{timeout-minutes} for all jobs, and pinning runner versions, which together correct over 50\% of the most common violations and narrow the gap between partially and fully compliant workflows. By focusing on these predictable structural gaps, developers can efficiently elevate workflow quality with low-effort interventions, creating a foundation for sustainable, maintainable CI configurations.
    
\section{Threats To Validity}
\label{sec:Threats_to_Validity}

    \subsection{Construct Validity}
    All constructs are derived from static YAML analysis, and therefore reflect how workflows are configured rather than how they behave at runtime. Complexity metrics capture configuration structure, not execution effort: a single \texttt{run:} step is treated uniformly regardless of whether it executes a trivial command or a large script, and YAML lines may overstate size for well-documented files. The external-action diversity ratio similarly assumes all steps are equivalent, which can distort the perceived extent of delegated automation. Semantic step classification relies on pattern matching over names and commands, making it sensitive to ambiguous or inconsistent naming. Measures used as proxies for quality or maturity also have limitations: the completeness score reflects the presence of canonical stages but does not account for modular designs that delegate functionality, while the absence of explicit test steps may underestimate testing if embedded within build tools. Trigger based measures capture declared configuration rather than actual usage, and the compliance score depends on author defined rule weights while covering only structural adherence. As a result, the constructs represent workflow structure rather than execution behavior or effectiveness.
   
    \subsection{Internal Validity}
    RQ1 performs 18 Kruskal-Wallis omnibus tests plus 54 pairwise Mann-Whitney follow-ups (three language pairs per metric) without a formal multiple-comparison correction. RQ2 and RQ3 apply the same significance threshold ($p < 0.05$) to eight continuous and twelve binary structural metrics (RQ2) and to compliance outcomes plus ten rule-violation indicators (RQ3), using Mann-Whitney $U$ and Fisher's exact tests, respectively. We report these tests primarily to confirm that cross-language differences are detectable at scale, but base our substantive conclusions on effect sizes (Cliff's $\delta$ for continuous metrics; Cohen's $h$ for proportions) rather than $p$-values alone, consistent with recommendations for large-sample software engineering studies where statistical significance is readily achieved but practical relevance must be assessed separately.
    
    \subsection{External Validity}
    RQ1 complexity findings use the full dataset of 27,863 workflows from 7,668 repositories, whereas the RQ2 and RQ3 findings use a stratified random sample of 382 workflows from the same dataset. Both units are drawn exclusively from public GitHub repositories in three language ecosystems (Python, Java, and C++), which bounds their generalizability. Proprietary and enterprise CI environments are not represented, and these contexts often exhibit different complexity patterns due to stricter compliance requirements, security policies, and standardized internal tooling. Even within open source, the dataset reflects a single temporal snapshot; as GitHub Actions evolves and newer features (e.g., reusable workflows) gain adoption, current patterns may not persist. The language scope further limits applicability, as other ecosystems such as JavaScript, Go, or Rust follow distinct build conventions and CI practices that may produce different structural characteristics. In addition, the per-language sample is uneven, with Python workflows dominating the dataset, which may skew aggregate observations, while smaller samples for Java and C++ limit the strength of conclusions for those groups. Finally, the stratified sampling frame defines the population to which the RQ2 and RQ3 results apply; any constraints in repository selection (e.g., activity level or prominence) directly bound the extent to which these findings can be generalized beyond the studied dataset.
    
\section{Conclusion}
\label{sec:Conclusion}
This study presents a large-scale empirical analysis of GHA usage across 27,863 workflows, with a manually examined subset of 382 workflows from Python, Java, and C++ open-source projects. Across structural, semantic, and compliance dimensions, workflows are predominantly small, shallow, and action-driven, with limited sequence-level standardization despite consistent intent-level patterns. Language is a statistically significant differentiator, reflecting ecosystem and tooling-specific conventions rather than intrinsic complexity. Systemic gaps are evident: many workflows omit explicit testing, secure practices such as SHA pinning and permission scoping are rarely implemented, and modularity mechanisms like reusable workflows are largely unused despite platform support. Compliance violations are widespread and systematic, driven by permissive defaults rather than isolated developer oversights.
These results indicate that CI practices in open-source repositories are shaped more by platform design, ecosystem norms, and toolchain constraints than by strict adherence to best practices. They highlight persistent structural, security, and modularity deficiencies while establishing a reproducible empirical baseline for monitoring GHA adoption. This baseline can inform targeted platform-level interventions, guide language-specific tooling improvements, and provide a foundation for future research aimed at improving CI workflow quality, security, and maintainability across diverse programming communities.

\medskip
\noindent{\textbf{Future Work.}} We aim to extend this work by studying the longitudinal evolution of CI workflows, analyzing repository-wide workflow interactions, and validating static complexity metrics using GitHub Actions runtime data. We also aim to expand the study to additional programming languages and CI platforms, account for repository characteristics, evaluate semantic compliance and supply-chain security practices, and investigate developer perspectives to better understand the motivations behind observed workflow patterns.

\section*{Declarations}

\bmhead{Funding} This research was supported by the Natural Sciences and Engineering Research Council (NSERC) Discovery Grant [RGPIN-2025-05897].

\bmhead{Acknowledgments}
The experiments conducted in this paper were enabled in part by support provided by the Digital Research Alliance of Canada. 

\bmhead{Author Contributions}{
Edward Abrokwah: Conceptualization, Investigation, Formal analysis, and Writing. 
Taher A. Ghaleb: Conceptualization, Investigation, Formal analysis, and Writing.
}

\bmhead{Data availability Statement}
The replication package for our experiments, including data, code, and results, is available on GitHub~\cite{our_replication_package}.

\bmhead{Conflicts of Interest} The authors declare that they have no known competing financial interests or personal relationships that could have appeared to influence the work reported in this paper.

\bmhead{Ethical approval} Not applicable.

\bmhead{Informed consent} Not applicable. 

\bmhead{Clinical Trial Number} Not applicable.

\bibliography{paper}

\end{document}